\documentclass[preprintnumbers,groupedaddress,nofootinbib,aps]{revtex4}

\usepackage{multirow}
\usepackage{amsmath,amssymb}
\usepackage{graphicx}
\usepackage{color}
\usepackage{url}
\usepackage{feynmp}
\usepackage[pdfborder={0 0 0}]{hyperref}
\hypersetup{
  pdfauthor={Malte Buschmann, Dorival Goncalves, Silvan Kuttimalai, Marek Schoenherr, Frank Krauss, Tilman Plehn},
  pdftitle={Top Mass Effects in the Higgs--Gluon Coupling: Boosted vs Off-Shell Production},
  pdfkeywords={QCD, NLO, Higgs Physics}
}
\begin{document}
\def\contentsname{{\normalsize Content}}
\def\tablename{Table}
\def\figurename{Figure}

\def\pveto{P_\text{veto}}
\def\nj{n_\text{jets}}
\def\meff{m_\text{eff}}
\def\ptmin{p_T^\text{min}}
\def\gtot{\Gamma_\text{tot}}
\def\as{\alpha_s}
\def\az{\alpha_0}
\def\gz{g_0}
\def\w{\vec{w}}
\def\sdag{\Sigma^{\dag}}
\def\s{\Sigma}
\newcommand{\psib}{\overline{\psi}}
\newcommand{\Psib}{\overline{\Psi}}
\newcommand\one{\leavevmode\hbox{\small1\normalsize\kern-.33em1}}
\newcommand{\Mpl}{M_\mathrm{Pl}}
\newcommand{\p}{\partial}
\newcommand{\mat}{\mathcal{M}}
\newcommand{\lag}{\mathcal{L}}
\newcommand{\ord}{\mathcal{O}}
\newcommand{\ope}{\mathcal{O}}
\newcommand{\qqquad}{\qquad \qquad}
\newcommand{\qqqquad}{\qquad \qquad \qquad}

\newcommand{\qb}{\bar{q}}
\newcommand{\matx}{|\mathcal{M}|^2}
\newcommand{\really}{\stackrel{!}{=}}
\newcommand{\msbar}{\overline{\text{MS}}}
\newcommand{\qns}{f_q^\text{NS}}
\newcommand{\lqcd}{\Lambda_\text{QCD}}
\newcommand{\met}{\slashchar{p}_T}
\newcommand{\pmiss}{\slashchar{\vec{p}}_T}

\newcommand{\sq}{\tilde{q}}
\newcommand{\go}{\tilde{g}}
\newcommand{\st}[1]{\tilde{t}_{#1}}
\newcommand{\stb}[1]{\tilde{t}_{#1}^*}
\newcommand{\nz}[1]{\tilde{\chi}_{#1}^0}
\newcommand{\cp}[1]{\tilde{\chi}_{#1}^+}
\newcommand{\cm}[1]{\tilde{\chi}_{#1}^-}
\newcommand{\CP}{CP}

% all the masses 
\providecommand{\mg}{m_{\tilde{g}}}
\providecommand{\mst}[1]{m_{\tilde{t}_{#1}}}
\newcommand{\msn}[1]{m_{\tilde{\nu}_{#1}}}
\newcommand{\mch}[1]{m_{\tilde{\chi}^+_{#1}}}
\newcommand{\mne}[1]{m_{\tilde{\chi}^0_{#1}}}
\newcommand{\msb}[1]{m_{\tilde{b}_{#1}}}
\newcommand{\vsm}{\ensuremath{v_{\rm SM}}}

% units of measure
\newcommand{\mev}{{\ensuremath\rm MeV}}
\newcommand{\gev}{{\ensuremath\rm GeV}}
\newcommand{\tev}{{\ensuremath\rm TeV}}
\newcommand{\fb}{{\ensuremath\rm fb}}
\newcommand{\ab}{{\ensuremath\rm ab}}
\newcommand{\pb}{{\ensuremath\rm pb}}
\newcommand{\sign}{{\ensuremath\rm sign}}
\newcommand{\iab}{{\ensuremath\rm ab^{-1}}}
\newcommand{\ifb}{{\ensuremath\rm fb^{-1}}}
\newcommand{\ipb}{{\ensuremath\rm pb^{-1}}}

% really great macro by Chris Lester
\def\slashchar#1{\setbox0=\hbox{$#1$}           % set a box for #1
   \dimen0=\wd0                                 % and get its size
   \setbox1=\hbox{/} \dimen1=\wd1               % get size of /
   \ifdim\dimen0>\dimen1                        % #1 is bigger
      \rlap{\hbox to \dimen0{\hfil/\hfil}}      % so center / in box
      #1                                        % and print #1
   \else                                        % / is bigger
      \rlap{\hbox to \dimen1{\hfil$#1$\hfil}}   % so center #1
      /                                         % and print /
   \fi}
\newcommand{\dslash}{\slashchar{\partial}}
\newcommand{\Dslash}{\slashchar{D}}

\newcommand{\eg}{\textsl{e.g.}\;}
\newcommand{\ie}{\textsl{i.e.}\;}
\newcommand{\etal}{\textsl{et al}\;}
%\DeclareMathOperator{\tr}{Tr}

% maximal number of floating environments on each page 
\setlength{\floatsep}{0pt}
\setcounter{topnumber}{1}
\setcounter{bottomnumber}{1}
\setcounter{totalnumber}{1}
\renewcommand{\topfraction}{1.0}
\renewcommand{\bottomfraction}{1.0}
\renewcommand{\textfraction}{0.0}
\renewcommand{\thefootnote}{\fnsymbol{footnote}}

\newcommand{\rig}{\rightarrow}
\newcommand{\lrig}{\longrightarrow}
\renewcommand{\d}{{\mathrm{d}}}
\newcommand{\be}{\begin{eqnarray*}}
\newcommand{\ee}{\end{eqnarray*}}
\newcommand{\gl}[1]{(\ref{#1})}
\newcommand{\ta}[2]{ \frac{ {\mathrm{d}} #1 } {{\mathrm{d}} #2}}
\newcommand{\bee}{\begin{eqnarray}}
\newcommand{\eee}{\end{eqnarray}}
\newcommand{\beeq}{\begin{equation}}
\newcommand{\eeeq}{\end{equation}}
\newcommand{\mc}{\mathcal}
\newcommand{\mr}{\mathrm}
\newcommand{\ep}{\varepsilon}
\newcommand{\emt}{$\times 10^{-3}$}
\newcommand{\emfo}{$\times 10^{-4}$}
\newcommand{\emfi}{$\times 10^{-5}$}

\newcommand{\revision}[1]{{\bf{}#1}}

\newcommand{\hzero}{h^0}
\newcommand{\Hzero}{H^0}
\newcommand{\Azero}{A^0}
\newcommand{\PHiggs}{H}
\newcommand{\PW}{W}
\newcommand{\PZ}{Z}

\newcommand{\sw}{\ensuremath{s_w}}
\newcommand{\cw}{\ensuremath{c_w}}
\newcommand{\swd}{\ensuremath{s^2_w}}
\newcommand{\cwd}{\ensuremath{c^2_w}}

%% 2HDM Higgs masses
\newcommand{\mhhd}{\ensuremath{m^2_{\Hzero}}}
\newcommand{\mhh}{\ensuremath{m_{\Hzero}}}
\newcommand{\mlhd}{\ensuremath{m^2_{\hzero}}}
\newcommand{\Mlh}{\ensuremath{m_{\hzero}}}
\newcommand{\mad}{\ensuremath{m^2_{\Azero}}}
\newcommand{\mhpd}{\ensuremath{m^2_{\PHiggs^{\pm}}}}
\newcommand{\mhp}{\ensuremath{m_{\PHiggs^{\pm}}}}

 \newcommand{\sa}{\ensuremath{\sin\alpha}}
 \newcommand{\ca}{\ensuremath{\cos\alpha}}
 \newcommand{\cad}{\ensuremath{\cos^2\alpha}}
 \newcommand{\sad}{\ensuremath{\sin^2\alpha}}
 \newcommand{\sbd}{\ensuremath{\sin^2\beta}}
 \newcommand{\cbd}{\ensuremath{\cos^2\beta}}
 \newcommand{\cb}{\ensuremath{\cos\beta}}
 \renewcommand{\sb}{\ensuremath{\sin\beta}}
 \newcommand{\tanbd}{\ensuremath{\tan^2\beta}}
 \newcommand{\cotbd}{\ensuremath{\cot^2\beta}}
 \newcommand{\tanb}{\ensuremath{\tan\beta}}
 \newcommand{\tb}{\ensuremath{\tan\beta}}
 \newcommand{\cotb}{\ensuremath{\cot\beta}}

\newcommand{\GeV}{\ensuremath{\rm GeV}}
\newcommand{\MeV}{\ensuremath{\rm MeV}}
\newcommand{\TeV}{\ensuremath{\rm TeV}}

\title{Mass Effects in the Higgs--Gluon Coupling: Boosted vs Off-Shell Production}

\author{Malte Buschmann$^{1,2}$, 
        Dorival Gon\c{c}alves$^2$, 
        Silvan Kuttimalai$^2$, 
        Marek Sch\"onherr$^2$,
        Frank Krauss$^2$, 
        Tilman Plehn$^1$}	

\affiliation{$^1$ Institut f\"ur Theoretische Physik, Universit\"at Heidelberg, Germany}
\affiliation{$^2$ Institute for Particle Physics Phenomenology, Department of Physics, Durham University, UK}

\begin{abstract}
  In the upcoming LHC run we will be able to probe the structure of
  the loop--induced Higgs--gluon coupling through kinematics. First,
  we establish state-of-the-art simulations with up to two jets to
  next-to-leading order including top mass effects. They allow us to
  search for deviations from the low-energy limits in boosted Higgs
  production. In addition, the size of the top mass effects suggests
  that they should generally be included in Higgs studies at the
  LHC. Next, we show how off-shell Higgs production with a decay to
  four leptons is sensitive to the same top mass effects. We compare
  the potential of both methods based on the same top--Higgs
  Lagrangian. Finally, we comment on related model assumptions
  required for a Higgs width measurement.
\end{abstract}

\maketitle

\bigskip \bigskip \bigskip
\tableofcontents

\newpage

%%%%%%%%%%%%%%%%%%%%%%%%%%
\section{Higgs--top sector at the LHC}
\label{sec:intro}

The recent discovery of a light, narrow, and likely fundamental Higgs
boson~\cite{higgs,discovery} makes studies of the properties of this
new particle one of the key tasks of the upcoming LHC run.  While the
Higgs coupling structure in the Higgs--gauge sector can be extracted
from precise tree--level information, our understanding of Higgs
couplings to fermions largely relies on loop--induced couplings.  This
is obvious when we look at our currently very limited and
model--dependent understanding of the top Yukawa
coupling~\cite{sfitter,couplings_ex,couplings_th,higgs_signals,higgs_review}. Its
measurement from associated Higgs and top pair production with a proper
reconstruction of the three heavy states is
challenging~\cite{tth,thj,th}.\bigskip

In generic models for physics beyond the Standard
Model~\cite{bsm_review} the effective gluon--gluon--Higgs vertex will receive
contributions from dimension--6 operators proportional to
$H^2G^{\mu\nu}G_{\mu\nu}$ and $H^2\bar{Q}_L\tilde{H}t_R$. One example
of new physics which can generate sizable perturbative corrections to
the Higgs--gluon coupling are light supersymmetric top
squarks~\cite{light_stop,schlaffer_spannowsky}.
%tp for BSM contributions decoupling does usually apply
%For these contributions the decoupling theorem does not apply, and thus 
%even in the limit of a large top--\-mass the coupling strength remains 
%finite and the corresponding coefficient does not vanish.  
Because of the non-decoupling structure of the Standard Model with
Yukawa couplings we can integrate out the top quark in the low-energy
limit, which describes the interactions between gluons and any Higgs
bosons in a simple effective Lagrangian~\cite{low_energy,lecture}. For
the top quark this effective Lagrangian provides a very good
prediction of the inclusive Higgs production rate with at most
$\ord(10\%)$ deviations in typical inclusive distributions for $gg \to
H$ production~\cite{spirix_nlo,spirix_review}.  On the other hand, the
same description fails spectacularly once the process becomes
sensitive to specific kinematic features, for instance for Higgs pair
production~\cite{higgs_pair}. Similarly, it fails for kinematic
observables which generate large momentum scales, such as off--shell
Higgs production with large invariant masses or the production of
Higgs bosons with large transverse momenta.  In these cases, the top
quark cannot be integrated out anymore, since the cross section does
not only include ratios of the kind $m_H/m_t$, but also ratios of
$Q/m_t$ with $Q$ being an additional hard scale.  In this paper we
consider effects of such additional large scales induced by the
observables, namely the production of single Higgs bosons at large
transverse momenta or far off--shell.

Being produced at large transverse momenta the Higgs boson will recoil
against a hard jet system.  For additional jets in the single Higgs
production process it is well known that the Higgs transverse momentum
distribution shows a logarithmic top mass dependence~\cite{keith,uli}.
Recently, this effect has been proposed as a handle to test the
structure of the loop--induced coupling and the underlying Standard
Model assumption that the Higgs--gluon coupling is exclusively due to
heavy quark loops~\cite{sanz,azatov,andi,robert,englert_spannowsky}.
In phase space regions where this logarithmic dependence occurs, the
two-jet contribution cannot be neglected~\cite{buschmann}.  This is
because it exhibits the same logarithmic structure as the one-jet
contribution and its rate at large transverse Higgs momenta is not
suppressed compared to the one-jet rate.  This necessitates an
accurate description of hard jet radiation beyond the simple parton
shower.\bigskip

An alternative method to probe the loop--induced Higgs--gluon coupling
is linked to off-shell production of the Higgs with a subsequent decay
into four leptons. Initially, it was noticed that off-shell Higgs
production and decay in this channel does not exhibit the usual
$\Gamma/m$ suppression~\cite{offshell,offshell_ww}. The reason for
this is that the off-shell suppression of the Higgs propagator is
partially compensated by lifting the off-shell suppression of the softer of the
$Z$ propagators. The interference with the $ZZ$ background further enhances
this effect. This problem can be turned into a virtue when we add
off-shell Higgs production with its modified dependence on the Higgs
width to the set of Higgs measurements~\cite{offshell_ex}. The initial
claim that this defines a `model independent' measurement of the Higgs
width ignores the fact that the Higgs--gluon coupling is induced by
loops and hence its momentum dependence only follows once we make an
assumption of the particles contributing to this
loop~\cite{offshell_model}. 

Alternatively, we can relate the width measurement to 
a determination of the full set of dimension-6 operators
and their coefficients induced by an unspecified new physics scenario~\cite{offshell_dim6}. 
Again, we can make use of a logarithmic top mass dependence, now in the distribution of the
momentum flowing through the Higgs propagator~\cite{glover_zz,taming}. While
this interpretation runs into problems with a consistent effective
theory description we can ask a slightly different question:
\textsl{can we track the top mass dependence of the loop--induced
  Higgs--gluon coupling and is there any indication for a more generic
  dimension-6 interaction?} This is exactly the same question which we
ask in boosted Higgs plus jets production, which means that we can
directly compare the potential of the two kinematic measurements. Moreover,
because we know that the top loop contributes to the effective Higgs--gluon
coupling we can link direct measurements of the top Yukawa to the 
study of the effective Higgs--gluon coupling and compare their respective
prospects~\cite{sfitter}.\bigskip

To link boosted Higgs production with off-shell Higgs production for
this specific physics question we define a theoretical framework. It
should allow us to test if the top Yukawa coupling is indeed
responsible for the observed Higgs--gluon coupling, or if other
particles contribute to the corresponding dimension-6 operator. The 
relevant part of the Higgs interaction Lagrangian including a finite top 
mass and free couplings reads~\cite{andi,buschmann}
\begin{alignat}{5}
\lag 
&= \lag_\text{SM} + 
\left[ \Delta_t \, g_{ggH} + \Delta_g \frac{\alpha_s}{12 \pi} \right] \; 
\frac{H}{v} \; 
G{\mu\nu}G^{\mu\nu}
- \Delta_t \; 
\frac{m_t}{v} H \left( \bar{t}_R t_L + \text{h.c.} \right) 
\qqquad &&\text{\textsc{SFitter}~\cite{sfitter}} \notag \\
&= \lag \Bigg|_{\kappa_j =0} + 
\left[  \kappa_t \; g_{ggH} + \kappa_g\frac{\alpha_s}{12 \pi}  \right] \; 
\frac{H}{v} \; 
G_{\mu\nu}G^{\mu\nu}
- \kappa_t \; 
\frac{m_t}{v} H \left( \bar{t}_R t_L + \text{h.c.} \right)
&&\text{Refs.~\cite{andi}} \; . 
\label{eq:lagrangian}
\end{alignat}
This Higgs--top Lagrangian will be the basis for the analysis
presented in this paper. The \textsc{SFitter} conventions show how
$\kappa_t$ as well as $\kappa_g$ are directly accessible in LHC
coupling analyses. While the effective coupling $g_{ggH}$ retains the
full top mass dependence, the dimension-6 operator is defined without
any reference to the top mass and assuming that the entire momentum
dependence arises from the gluon field strengths. We will use the
\textsc{SFitter} interpretation to eventually compare the expected
performance of the distribution--based search strategies to the usual
Higgs coupling analysis. The Standard Model limit is given by
$\Delta_g = 0 = \Delta_t$.
%One physics scenario which could serve as an
%ultraviolet extension of Eq.\eqref{eq:lagrangian} would be the
%Standard Model with an extended Higgs sector and an unobserved top
%partner~\cite{sfitter,schlaffer_spannowsky}. 
Our two reference points
will be
\begin{alignat}{5}
(\kappa_t,\kappa_g)_\text{SM}=(1,0) 
\qqquad \text{and} \qqquad 
(\kappa_t,\kappa_g)_\text{BSM}=(0.7,0.3) \; .
\label{eq:points}
\end{alignat}
In the second point the contributions from a top partner to a good
approximation compensate for the reduced top Yukawa in the
Higgs--gluon coupling, leaving the observed Higgs cross section at the
LHC unchanged. This last condition is crucial to get a realistic
estimate of the power of distribution--based Higgs analyses, because a
significant deviation of the Higgs production rate in gluon fusion
will be experimentally accessible long before boosted or off-shell
Higgs analysis will become sensitive.

%%%%%%%%%%%%%%%%%%%%%%%%%%%%%%%%%%%%%%%%%%%%%%%%%%
\section{Top mass effects in Higgs rates}
\label{sec:mass}

As long as we limit ourselves to the total cross section of the Standard
Model Higgs boson, the heavy top limit or low-energy limit provides an
accurate prediction for the total rate. The effective Higgs--gluon coupling
is then given by a single coupling value~\cite{low_energy,lecture}
\begin{alignat}{5}
\lag_{ggH} &\supset \;
g_{ggH} \; \frac{H}{v} \;  \, G^{\mu\nu}G_{\mu\nu} \notag \\
\frac{g_{ggH}}{v} 
 &=
\,\frac{\alpha_s}{8 \pi} \; 
   \frac{1}{v} \; \tau \left[1+(1-\tau)f(\tau)\right] 
\qqqquad 
f(\tau) 
\stackrel{\text{on-shell}}{=} 
\left( \arcsin \sqrt{\dfrac{1}{\tau}}\right)^2 
\stackrel{\tau \to \infty}{=} 
\frac{1}{\tau} + \frac{1}{3\tau^2} + \ope \left( \frac{1}{\tau^3} \right) 
\; ,
\label{eq:higgs_eff1}
\end{alignat}
all in terms of $\tau = 4m_t^2/m_H^2 > 1$.  The usual scalar
three-point function is written in the dimensionless form $f(\tau) = -
m_H^2 C(m_H^2;m_t,m_t,m_t)/2$. The effective coupling $g_{ggH}$ of an
on-shell Higgs boson to incoming or outgoing gluons accounts for all
top mass effects in the three--particle Green's function. However,
whenever we go beyond the assumption of three on-shell particles it
fails to describe the dynamics. For example, applied to differential
distributions including additional jets, like the transverse momentum
of the Higgs boson in LHC production, the low-energy approximation in
the dimension-6 operator breaks down.  In this regime, the top
contribution in the loop starts to be resolved and leads to effects in
the distributions. Similarly, when we allow the singly produced Higgs
boson to be off-shell, the effective Higgs--gluon coupling $g_{ggH}$
becomes a non-trivial function of the mass scales involved.\bigskip

%-------------------------------------------------------
\begin{figure}[b!]
\includegraphics[width=0.95\textwidth]{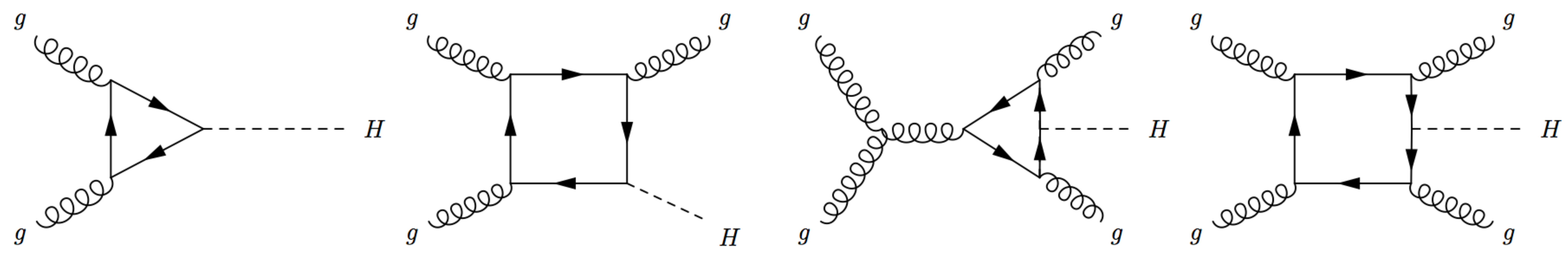}
\caption{Sample Feynman diagrams contributing for Higgs production
with up to 2 jets at leading order.}
\label{fig:born_diag} 
\end{figure}
%-------------------------------------------------------

In this section we present a state--of--art event simulation including
top mass effects beyond the low-energy limit.  It relies on the
general purpose Monte Carlo event generator \textsc{Sherpa}~\cite{sherpa}.
We use \textsc{Sherpa} to generate events for Higgs boson production in
association with up to 2 jets and the corresponding backgrounds, $WW$ and 
top pair production.  In all cases we apply multi-jet merging of the matrix elements
and the parton showers at LO with the algorithm presented in~\cite{ckkw,mets} 
or, where not otherwise stated at NLO according to the \textsc{Meps@Nlo}
algorithm~\cite{mepsnlo}.  In each case, the implementation is automated 
once the respective virtual matrix elements are available.  This way we 
generate NLO-merged events for Higgs production with jets in the low-energy
limit~\cite{sherpa_h}. In this paper we extend this implementation by
top (and bottom) mass effects at leading order accuracy by using loop--\-level
matrix elements provided by \textsc{OpenLoops}~\cite{openloops}, which we
use for reweighting the effective theory.\bigskip

%-------------------------------------------------------
\begin{figure}[t]
\includegraphics[width=0.48\textwidth]{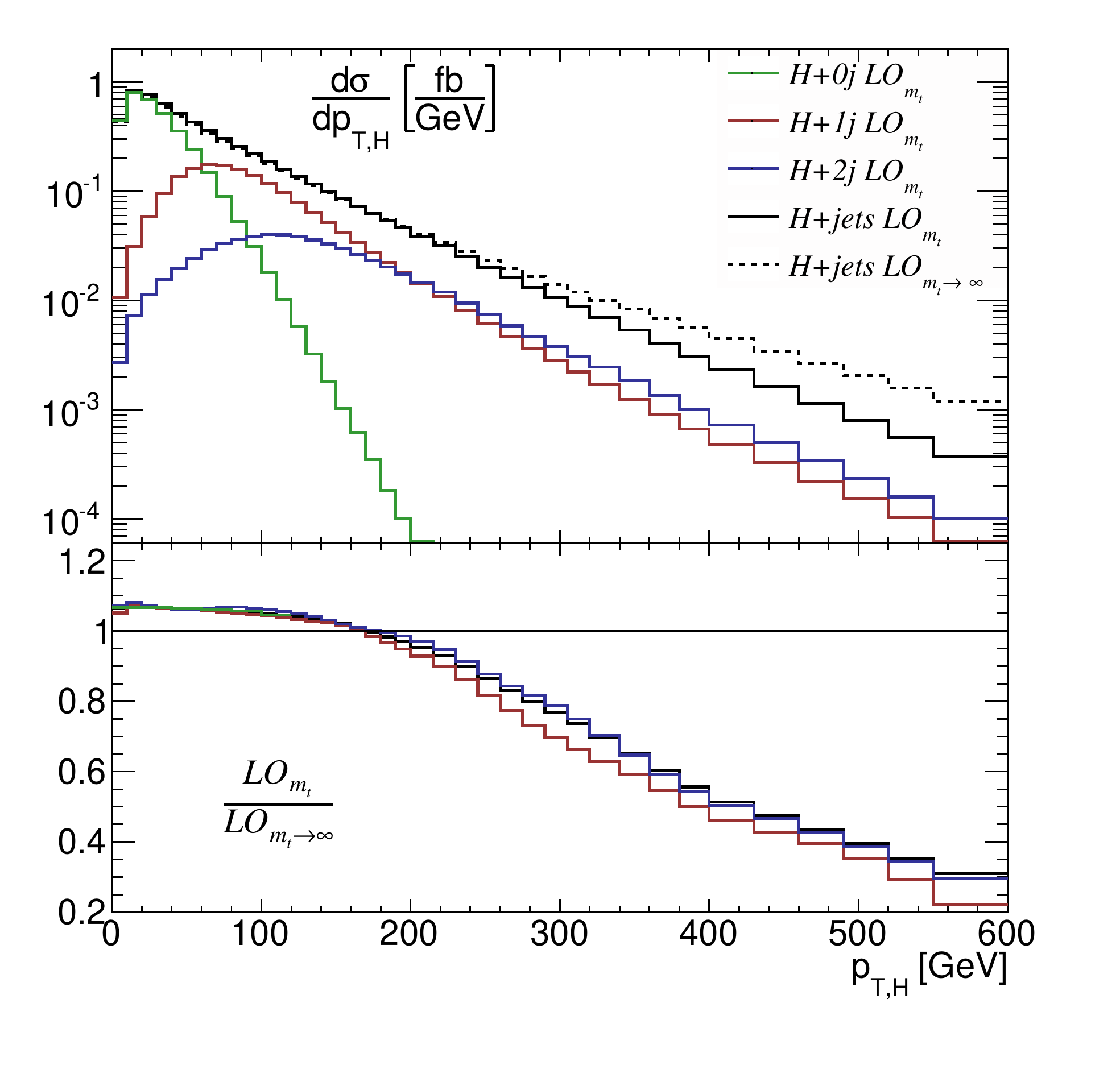}
\includegraphics[width=0.48\textwidth]{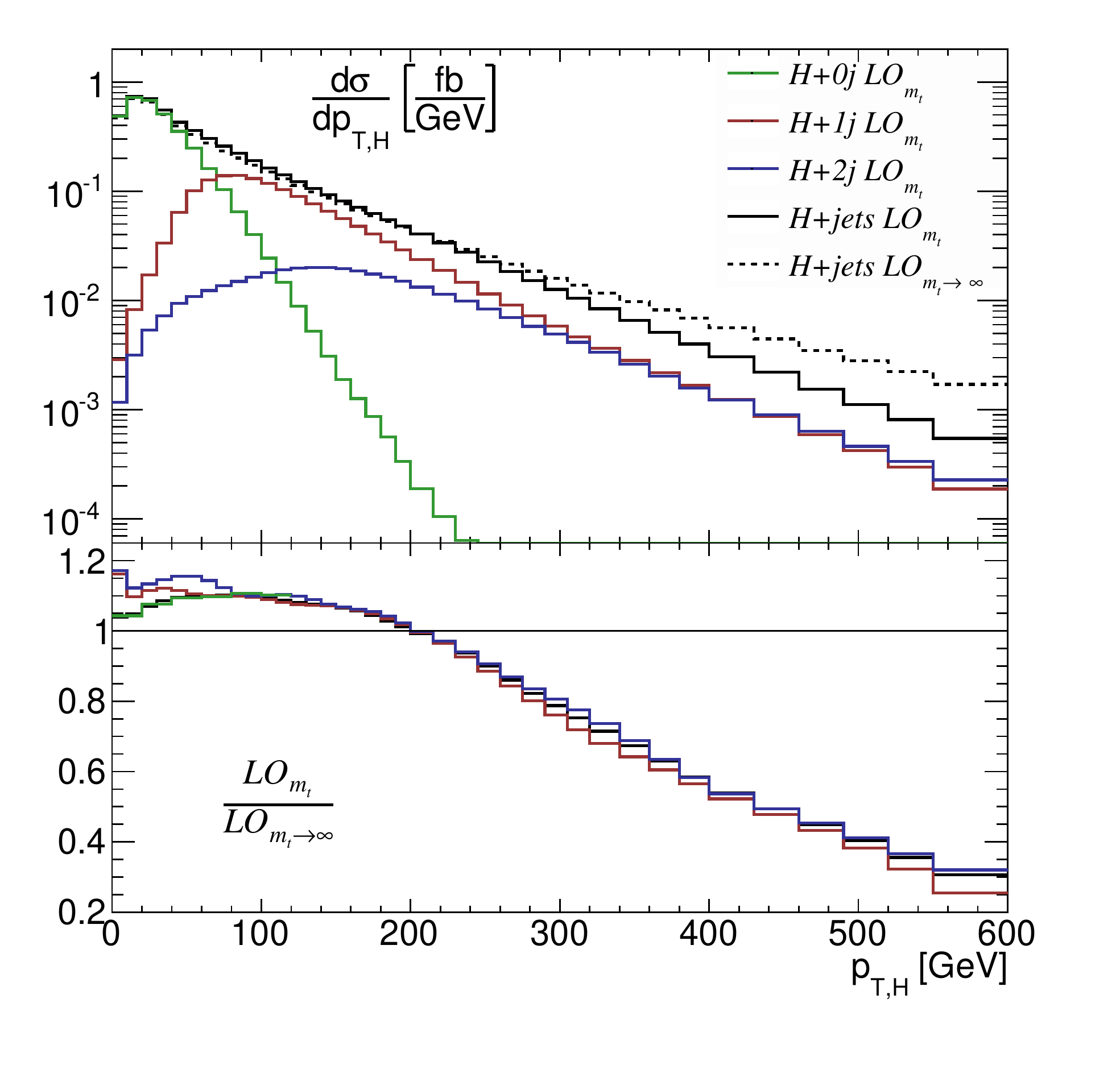}
\vspace{-0.6cm}
\caption{Transverse momentum distribution $p_{T,H}$ for 
$H\rightarrow WW$+jets production at LO with \textsc{Sherpa} (left panel) 
and \textsc{Pythia8} (right panel). We present the distributions for
  exclusive and merged jet samples with finite top mass effects 
  $(m_t=173~\gev)$ and in the low-energy approximation 
  $(m_t\rightarrow \infty)$. We assume the LHC at $\sqrt{S}=13$~TeV.}
\label{fig:pth_merged_LO} 
\end{figure}
%-------------------------------------------------------
 
As a starting point, we study the impact of the top mass corrections
in the Higgs boson transverse momentum using LO multi-jet merging with
up to two jets.  A sample of the representative Feynman diagrams are
displayed in Fig.~\ref{fig:born_diag}.  We assume fully leptonic Higgs
decays $H\rightarrow W_\ell W_\ell$+jets. In the left panel of
Fig.~\ref{fig:pth_merged_LO} we display the LO-merged results from
\textsc{Sherpa}. Because this automized implementation relies on the
low-energy limit for the Higgs--gluon coupling described in
Eq.\eqref{eq:higgs_eff1}, we reweight all tree-level matrix elements
in the effective theory with their full loop counterparts. This
defines a correction factor~\cite{fabio}
\begin{alignat}{5}
r^{(n)}_t = 
\frac{|\mathcal{M}^{(n)}(m_t)|^2}
     {|\mathcal{M}^{(n)}(m_t \to \infty)|^2}
\label{eq:r_t}
\end{alignat}
for each jet multiplicity $n$. In the right panel of
Fig.~\ref{fig:pth_merged_LO} we also show the corresponding results
from \textsc{Pythia8}~\cite{pythia}, based on the CKKW-L merging. The
parton level events for the \textsc{Pythia} merging in the 0-jet bin
come from \textsc{MadGraph5}~\cite{mg5}, in the 1-jet bin we use
\textsc{MCFM}~\cite{mcfm}, and in the 2-jet bin we use
\textsc{VBFNLO}~\cite{vbfnlo}.

At the analysis level, jets are defined using the anti-kT algorithm 
implemented in Fastjet  with R = 0.5 and we assume basic acceptance cuts
\begin{alignat}{5}
p_{T,\ell}&>20~\gev \qquad & |\eta_\ell|&<2.5 \notag\\
p_{T,j}&>30~\gev \qquad & |\eta_j|&<4.5 \;.
\label{eq:boosted_cuts}
\end{alignat}
The individual curves for the different jet bins account for the
number of jets passing these acceptance cuts, rather than the number
of hard jets entering the merging procedure. The results from
\textsc{Sherpa} and \textsc{Pythia8} broadly agree with each other. In both
simulations we observe that for all contributions the low-energy limit
and the full results scale the same way as long as $p_{T,H} \lesssim
m_t$.  In this regime the only difference is a constant scaling factor
1.065 for the Higgs--gluon coupling. Although  $b$-quark loops become relevant in this regime and
need to be accounted, they present sub-leading contributions 
in the boosted regime. Since we will be mostly concerned with boosted Higgs of
$p_{T,H} \gtrsim m_t$ these contributions can be safely 
neglected~\cite{bottom_effects}.

Above this energy scale the effective and full theory start to visibly 
diverge. Looking at the jet multiplicities we confirm that this effect is driven by Higgs
production with two jets, where the top mass effects are not only
larger than in the one-jet process relative to the respective cross
section, but also larger in absolute terms~\cite{buschmann}. In the
lower panels of Fig.~\ref{fig:pth_merged_LO} we see that the top mass
effects lead to contributions as large as a factor four in the rate at
transverse momenta of $600~\gev$\footnote{The size of these effects
  suggest that for \textsl{any} strongly boosted Higgs analysis a
  proper modelling of the top mass effects is of vital importance.}.
Another remarkable feature which we observe in
Fig.~\ref{fig:pth_merged_LO} is that the top mass effects factorize:
the full top mass dependence provides the same $p_{T,H}$-dependent
correction factor for each jet bin, and consequently for the merged
result. Finally, the lower panels of Fig~\ref{fig:pth_merged_LO}
indicate that an experimental analysis including systematic and
theoretical uncertainties can rely on the range $p_{T,H} < m_t$ as a
safe reference region, searching for a distinct turn-over in the
distribution around $p_{T,H} = m_t$.\bigskip

%-------------------------------------------------------
\begin{figure}[t!]
\includegraphics[width=0.95\textwidth]{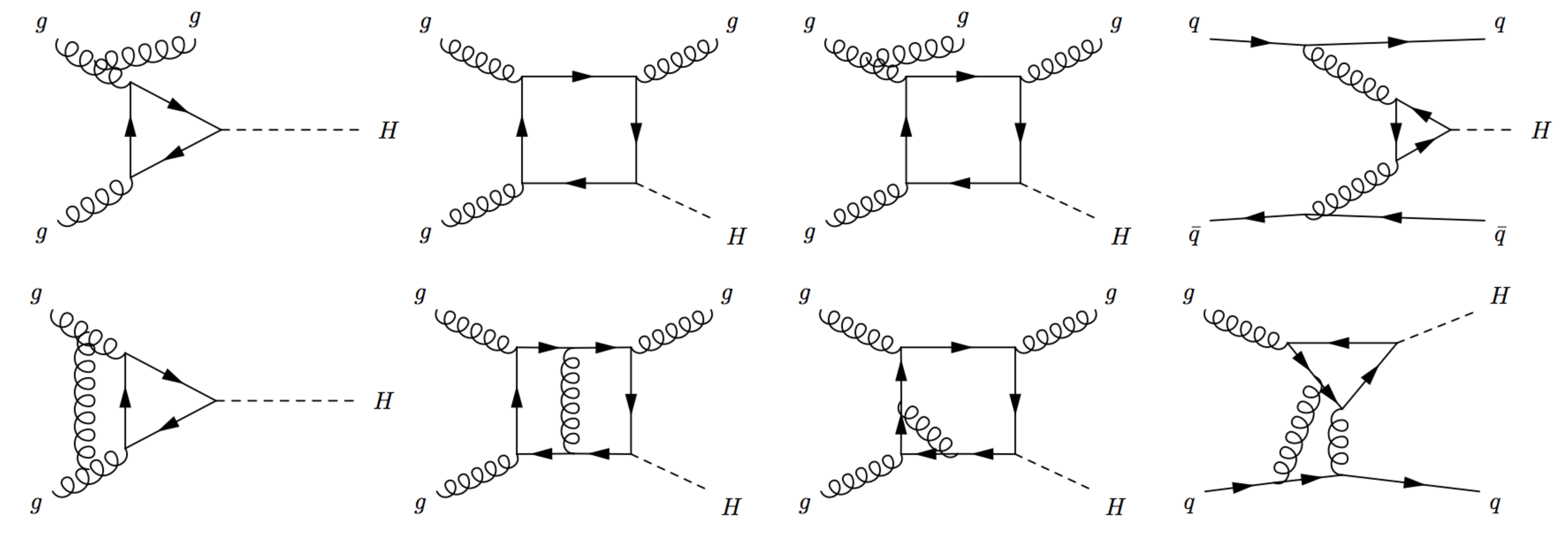}
\caption{Sample one-loop Feynman diagrams contributing to the Higgs and Higgs-jet production.
On the top we display the NLO real corrections and on the bottom the virtual contributions.}
\label{fig:nlo_diag} 
\end{figure}
%-------------------------------------------------------

This observed factorization at leading order strengthens the basic
assumption underlying our precision study, namely that top mass
effects in Higgs production are fully associated with the hard
process. This is know to not apply to bottom mass effect, which we
assume to be small and not critical for the phase space regions we
consider~\cite{bottom_effects}.  Hence, we can use the
\textsc{Sherpa} results in the low-energy limit and reweight them on
an event-by-event basis with the corresponding heavy-quark matrix
element. 

The \textsc{Meps@Nlo} algorithm~\cite{mepsnlo} for multi-jet merging of
NLO matrix elements can be viewed, intuitively, as stacking towers of
individual \textsc{Mc@Nlo} simulations~\cite{mcatnlo} on top of each other, 
without a double counting of emissions.  The only subtlety in the
\textsc{Sherpa} implementation~\cite{smcatnlo} is that the actual 
implementation of the \textsc{Mc@Nlo} algorithm has been slightly changed to 
also include sub--\-leading color effects in the Sudakov form factor.  
To see in more detail how this works at NLO, let us consider the structure 
of the \textsc{S-Mc@Nlo} cross section (including the first emission)
\begin{alignat}{5}
  \text{d}\sigma^\text{\textsc{S-Mc@Nlo}}=\;
  &\text{d}\Phi_n \left[\mathcal{B}+\mathcal{V}
                          +\int\text{d}\Phi_1\;\mathcal{D}
                    \right]
   \left(\Delta(t_0)
         +\int\text{d}\Phi_1\,
              \frac{\mathcal{D}}{\mathcal{B}}\,\Delta(t)\right)
 +
   \text{d}\Phi_{n+1}\big[\mathcal{R}-\mathcal{D}\big]\; ,
  \label{eq:nlo1}
\end{alignat}
where $\mathcal{B}$, $\mathcal{V}$ and $\mathcal{R}$ denote the Born,
virtual and real emission contributions associated with the $n$ and
$n+1$ particle phase space integrals.  The \textsc{S-Mc@Nlo}
resummation kernel
$\mathcal{D}=\widetilde{\mathcal{B}}\otimes\mathcal{K}$ is constructed
from a color-correlated and spin-correlated Born matrix element
$\widetilde{\mathcal{B}}$ and a suitable splitting function
$\mathcal{K}$~\cite{mcatnlo,csdipole}. By construction, $\mathcal{D}$
coincides with the real emission matrix element in the soft and/or
collinear limit. Note that in \textsc{S-Mc@Nlo} the ratio 
$\mathcal{D}/\mathcal{B}$ also constitutes the kernel of the Sudakov form 
factor for the first emission, in difference to the original \textsc{Mc@Nlo}
method.

In a second step we reweight all tree-level matrix
elements in the low-energy limit with their full loop
counterparts. This gives rise to correction factors $r^{(n)}_t$
defined in Eq.\eqref{eq:r_t} modifying the merged rate prediction in
Eq.\eqref{eq:nlo1},
\begin{alignat}{5}
  \text{d}\sigma^\text{\textsc{S-Mc@Nlo}}=\;
  &\text{d}\Phi_n\,r^{(n)}_t\,
                    \left[\mathcal{B}+\mathcal{V}
                          +\int\text{d}\Phi_1\;\mathcal{D}
                    \right]
   \left(\Delta(t_0)
         +\int\text{d}\Phi_1\,
              \frac{\mathcal{D}}{\mathcal{B}}\,\Delta(t)\right)
  +
   \text{d}\Phi_{n+1}\left[r^{(n+1)}_t\mathcal{R}
                             -r^{(n)}_t\mathcal{D}
                       \right]\; .
  \label{eq:nlo2}
\end{alignat}
The NLO corrections in the low-energy limit and the top mass
corrections are thus applied in a factorized form. This prescription
offers a gauge invariant interpolation between both types of
corrections. It is worth noting that the resummation properties of the
\textsc{S-Mc@Nlo} kernel are not altered, because its argument is a
ratio of matrix elements. The infrared safety of the fixed-order
correction is guaranteed as long as $r_t^{(n+1)}\to r_t^{(n)}$ in the
infrared limit. Our approach generalizes the \textsc{Meps@Nlo}
method~\cite{mepsnlo}, now including next-to-leading order corrections
in the low-energy approximation as well as the top mass dependence at
leading order for all jet multiplicities considered.  Eventually, it
needs to be tested once the two-loop multi-scale diagrams can be
evaluated over the full phase space.\bigskip

\begin{figure}[t!]
\includegraphics[width=0.48\textwidth]{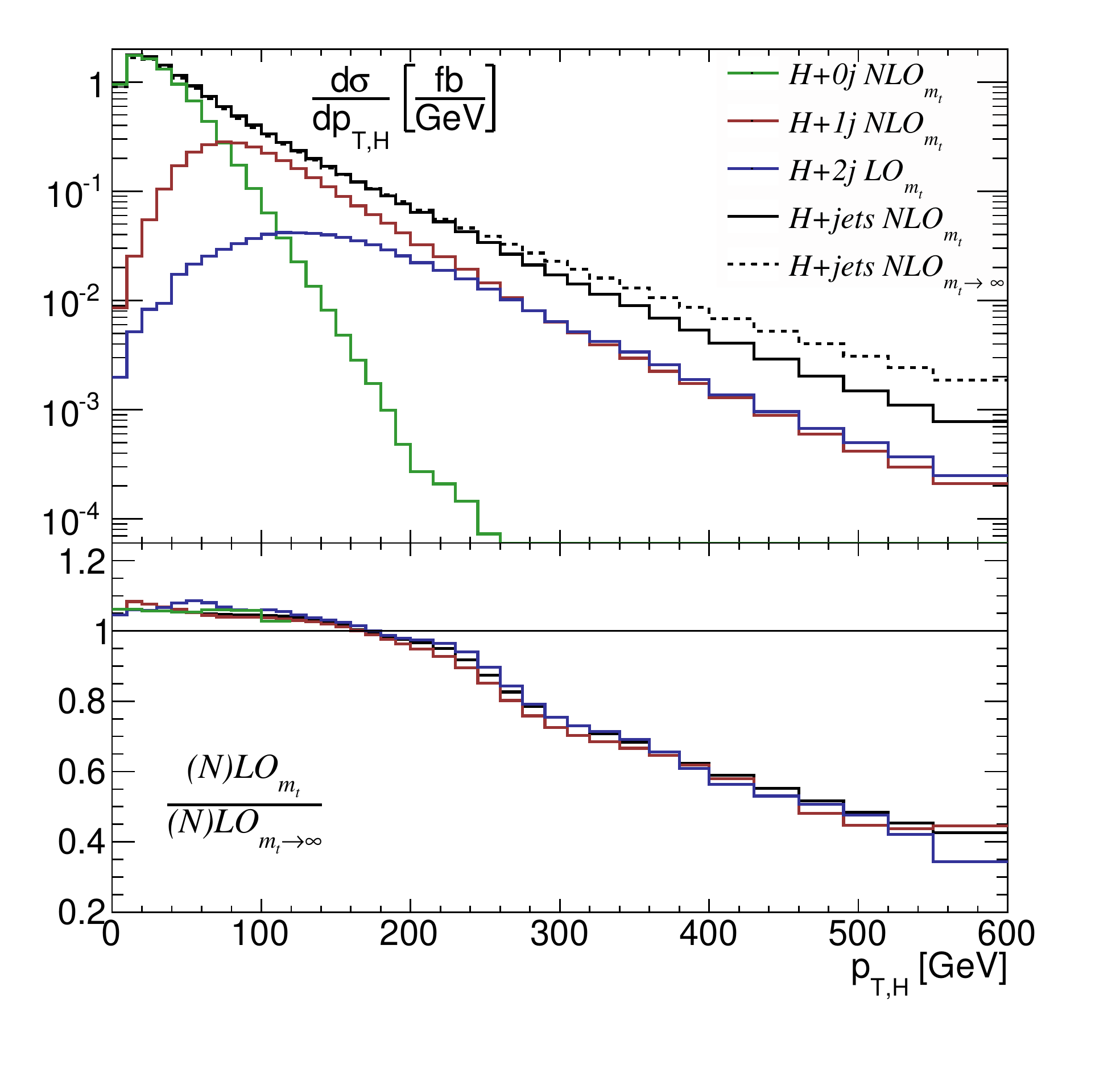}
\includegraphics[width=0.48\textwidth]{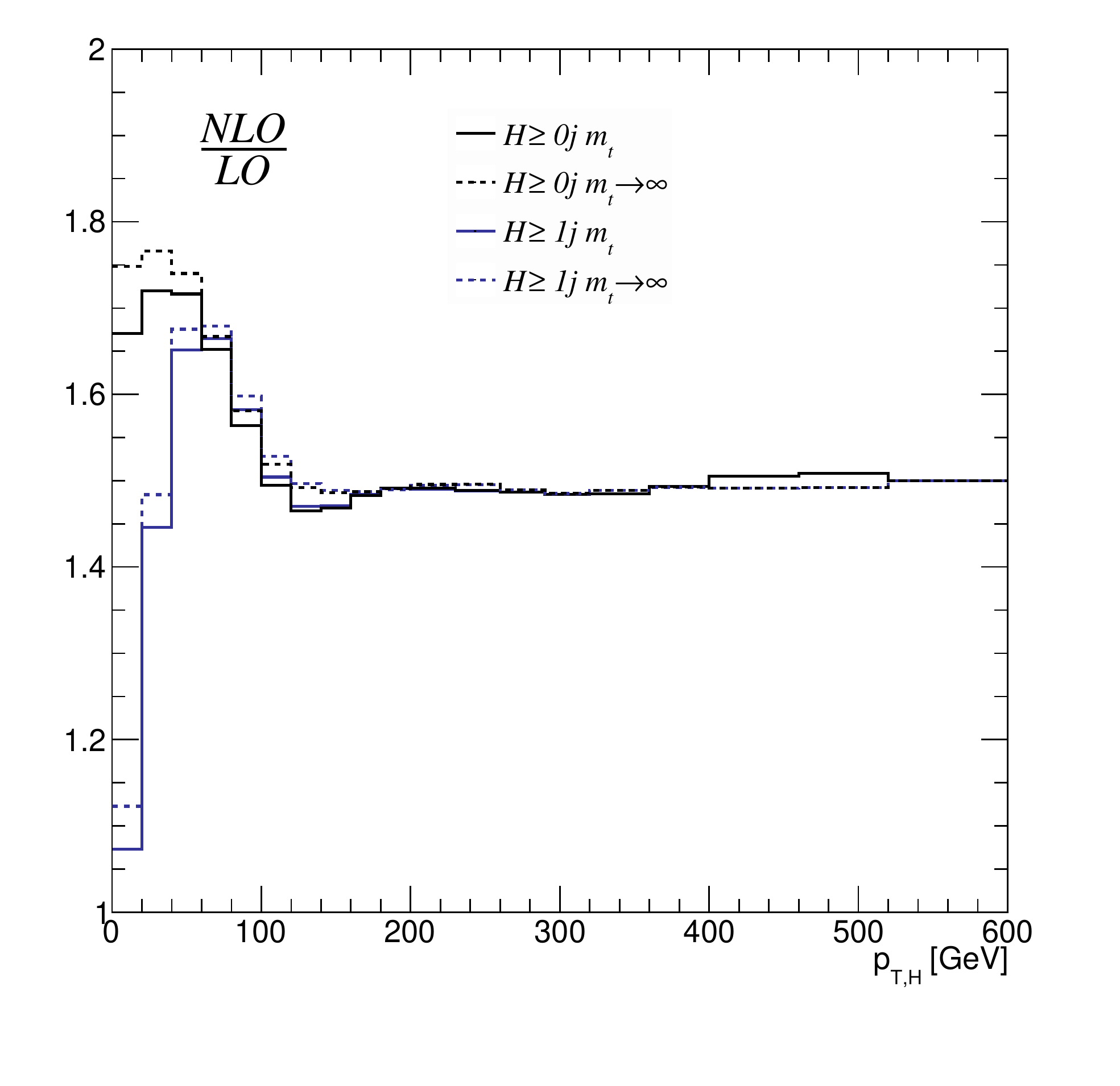}
\vspace{-0.6cm}
\caption{Transverse momentum distribution $p_{T,H}$ for $H\rightarrow
  WW$+jets production with \textsc{Sherpa} at NLO (left panel). We
  present the distributions for exclusive and merged jet samples with
  finite top mass effects $(m_t= 173~\gev)$ and in the low-energy
  approximation $(m_t\rightarrow \infty)$. In the right panel we show
  the $p_{T,H}$-dependent $K$-factor for $H$ and $H+1$~jet production.}
\label{fig:pth_merged_NLO} 
\end{figure}

Following this implementation we upgrade our boosted Higgs analysis in
Fig.~\ref{fig:pth_merged_LO} to the NLO level. The 0-jet and 1-jet
bins include the NLO corrections, while the 2-jet bin remains at
leading order. In Fig.~\ref{fig:nlo_diag} we display a sample of the
Feynman diagrams in the NLO corrections.  The upgraded NLO
distributions are presented in Fig.~\ref{fig:pth_merged_NLO}.  In the
left panel we show that apart from the different total rate all top
mass features are completely analogous to the leading order case.  The
ratio between the full calculation and low-energy limit shows the same
profile. In the right panel, we shown that the NLO corrections
factorize, \ie the relative NLO corrections for the full theory and
for the low-energy approximation agree independently for the $H$ and
$H+1$~jet rate.

%-------------------------------------------------------
%-------------------------------------------------------

%%%%%%%%%%%%%%%%%%%%%%%%%%%%%%%%%%%%%%%%%%%%%%%%%%
\section{Boosted Higgs production}
\label{sec:boost}

For boosted Higgs production the effect of a finite top mass has been known
for an eternity: adding jets to the hard process pushes one or two
gluon propagators off their respective mass shell. In that case the
matrix elements for Higgs production in association with one
jet~\cite{keith,uli,sanz} and two jets~\cite{buschmann} develop a top mass
dependence,
\begin{alignat}{5}
|\mat_{Hj(j)}|^2 
\propto 
m_t^4 \; \log^4 \frac{p_{T,H}^2}{m_t^2} \; .
\label{eq:pt_log}
\end{alignat}
Beyond this logarithmic dependence absorptive parts of the one-loop
integrals exist, but are unfortunately too small to be observed in the
coming LHC run(s)~\cite{buschmann}.  If we follow
Eq.\eqref{eq:lagrangian} and allow for a top quark as well as an
unspecified heavy state in the loop-induced Higgs--gluon
coupling, we can write the matrix element for Higgs production in gluon
fusion as
\begin{alignat}{5}
\mat = \kappa_t \mat_t + \kappa_g \mat_g \; .
\label{eq:amplitude_boost}
\end{alignat}
The index $t$ marks the top contributions, while 
$g$ the contributions from the dimension-6 Higgs--gluon operator in the low-energy limit.
All prefactors except for the $\kappa_j$ are absorbed in the definitions of $\mat_j$.\bigskip

%-------------------------------------------------------
\begin{figure}[t]
 \includegraphics[width=0.38\textwidth]{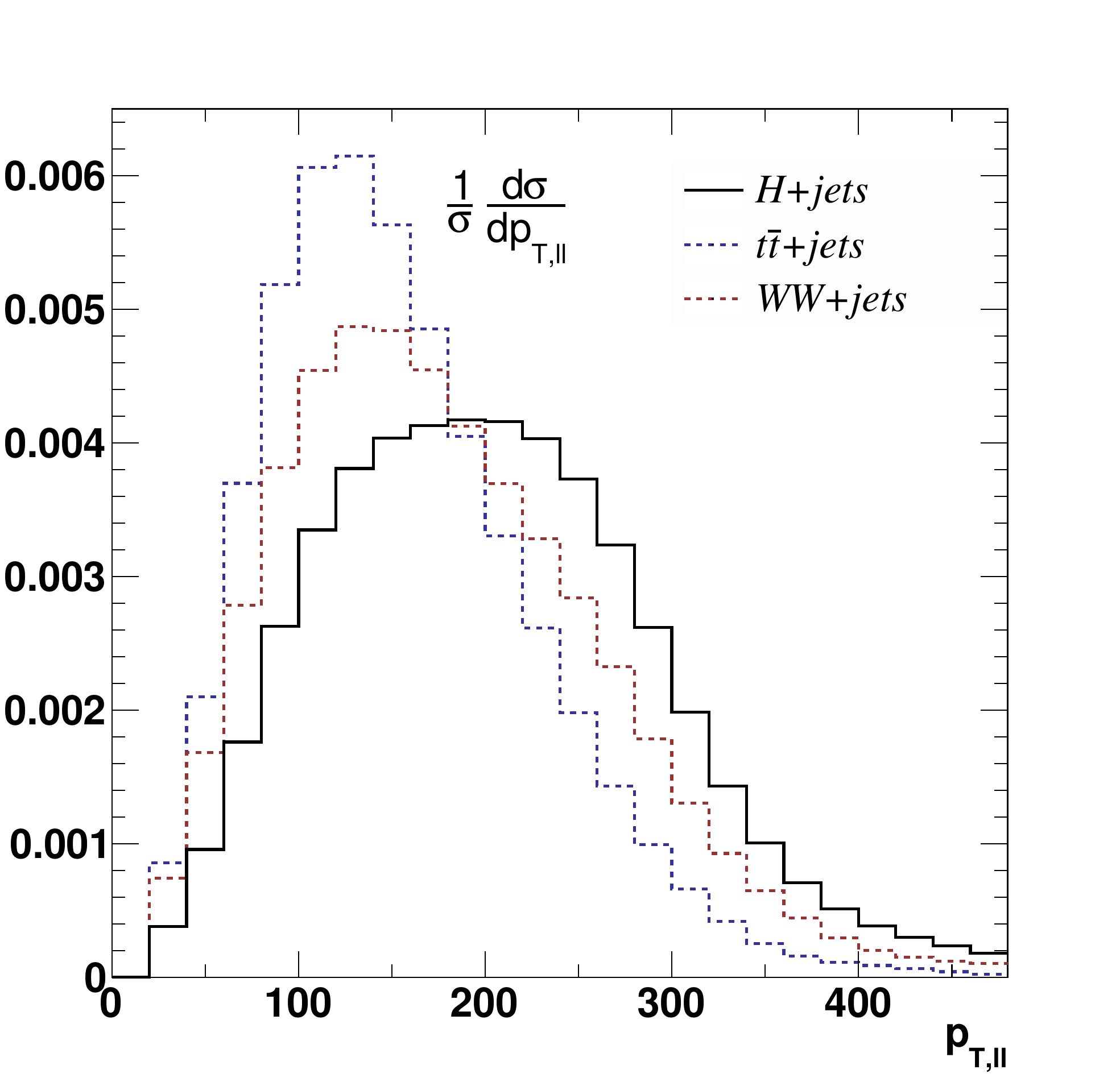}
 \hspace*{0.1\textwidth}
 \includegraphics[width=0.38\textwidth]{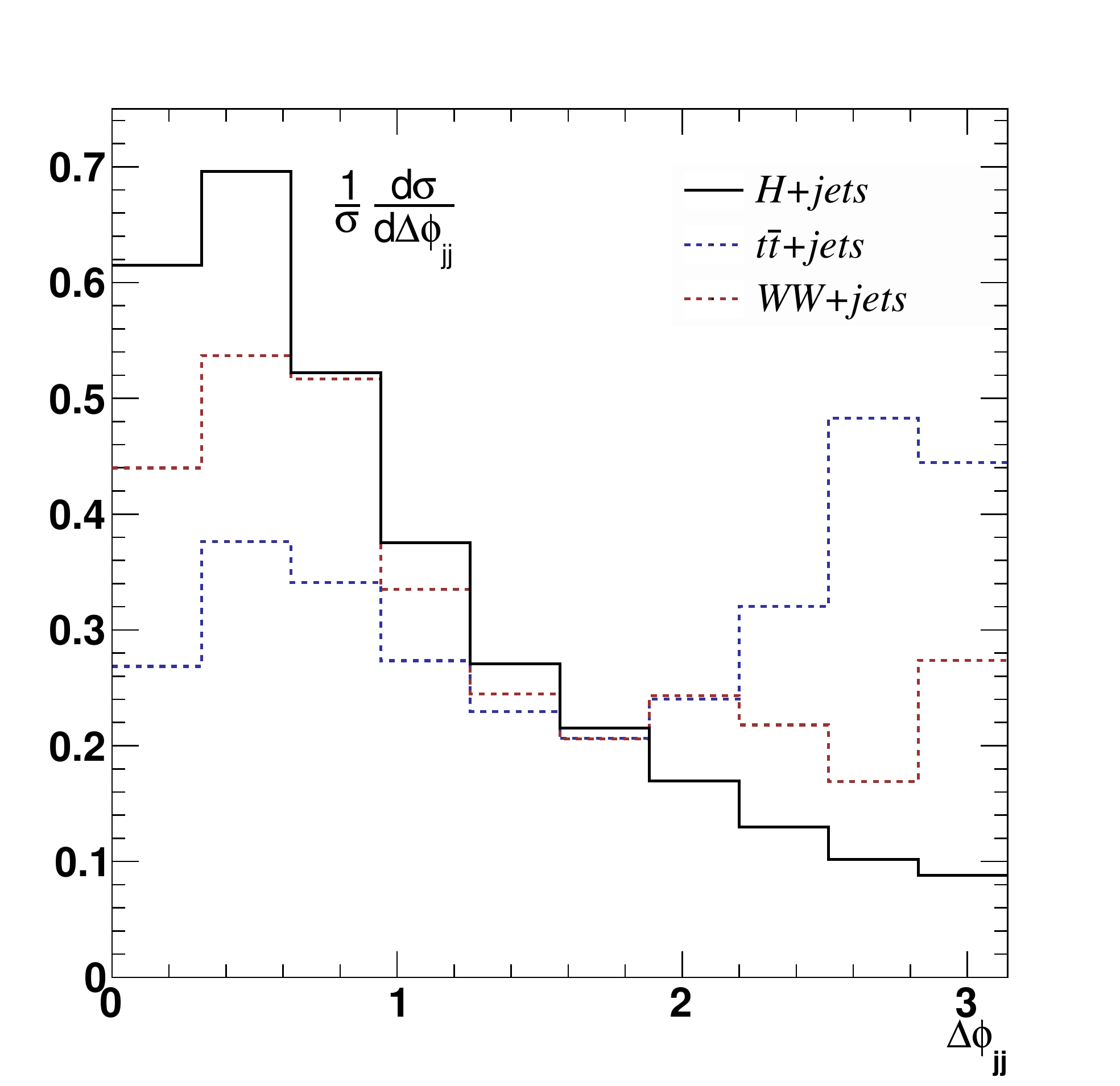}
\caption{Normalized $p_{T,ll}$  (left) and $\Delta\phi_{jj}$ (right) 
distributions for the $H\rightarrow WW$ signal and the
  dominant backgrounds. All universal cuts listed in
  Tab~\ref{tab:cuts1} are already applied.
  We assume the LHC at $\sqrt{S}=13$~TeV.}
\label{fig:cuts1} 
\end{figure}
%-------------------------------------------------------

%-------------------------------------------------------
\begin{table}[b!]
\begin{tabular}{l || c | c | c || c | c | c  }
  \multicolumn{1}{c||}{} &
  \multicolumn{3}{c||}{$(H \to WW)+ (0+1)j$}&
  \multicolumn{3}{c}{$(H \to WW)jj$ inclusive}
    \\
  \hline
 \multirow{1}{*}{cuts} &  
 \multicolumn{1}{c|}{$H$+jets} & 
 \multicolumn{1}{c|}{$WW$+jets}  & 
 \multicolumn{1}{c||}{$t\bar{t}$+jets}&
 \multicolumn{1}{c|}{$H$+jets} & 
 \multicolumn{1}{c|}{$WW$+jets}  & 
  \multicolumn{1}{c}{$t\bar{t}$+jets} \\ 
  \hline
{$p_{T,j}>40$~\gev, $|y_j|<4.5$}
&\multirow{2}{*}{87.9 } & \multirow{2}{*}{3220} &\multirow{2}{*}{9640 } 
&\multirow{2}{*}{ 6.50} & \multirow{2}{*}{203} &\multirow{2}{*}{5890}\\  
$p_{T,\ell}>20$~\gev, $|y_\ell|<2.5$
 &  & & & &  & \\  
$N_b=0$                
& 84.9 & 3180 & 7400
& 5.09 & 189 & 2790  \\ 
$m_{\ell \ell}\in [10,60]~\gev$               
& 69.0 & 628 & 1470  
& 4.22 & 36.2 & 503  \\ 
$\slashchar{E}_T>45~\gev$               
& 49.7 & 504 & 1250  
& 3.55 & 32.6 & 493  \\
$\Delta \phi_{\ell \ell}<0.8$               
& 24.0 & 195 & 561  
& 2.78 & 20.2 & 237   \\
$m_T<125~\gev$               
& 23.7 & 74.5 & 250  
& 2.75 & 10.8 & 119   \\
$p_{T,H}>300~\gev$               
& 0.27  & 1.41 & 1.24  
& 0.42 & 2.12 & 5.32 \\
$p_{T,ll}>180~\gev$               
& 0.15 & 0.58 & 0.35   
& 0.24 & 0.98 & 1.87 \\ \hline
$\Delta \phi_{jj}<1.8$               
&  &&
& 0.21 & 0.69 & 0.90  
\end{tabular} 
\caption{Cut flow for $H$+jets, $WW$+jets and $t\bar{t}$+jets.  All
  events are generated with Sherpa using the MEPS@NLO algorithm. The
  rates are given in fb.}
\label{tab:cuts1}
\end{table}
%-------------------------------------------------------

Following the bench mark point in Eq.\eqref{eq:points} we will be
specially interested in deviations from the Standard Model, where the
two couplings satisfy ${\kappa_t+\kappa_g=1}$. The transverse momentum
distribution will then allow us to disentangle the effects of
$\kappa_t$ and $\kappa_g$ while respecting the experimental constrains
on the Higgs production cross section ${\sigma \sim
  |\kappa_t+\kappa_g|^2}$.
For a kinematic distribution like the Higgs transverse momentum this means
\begin{alignat}{5}
\frac{d\sigma}{dp_{T,H}} & 
= \kappa_t^2 \; \frac{d\sigma_{tt}}{dp_{T,H}}  
+ \kappa_t \kappa_g \; \frac{d\sigma_{tg}}{dp_{T,H}}  
+ \kappa_g^2 \; \frac{d\sigma_{gg}}{dp_{T,H}} \; .
\label{eq:distribution_boost}
\end{alignat}
To access the different components in the Higgs--gluon coupling one
needs to decouple the soft and hard momentum components flowing in
this loop--induced coupling.  The separation of these factors can be
efficiently achieved through the kinematics of Higgs plus jets
production.  This feature was studied for the 1-jet or 2-jet cases,
indicating that we can achieve a decent sensitivity for ratio for
integrated luminosities of $\ope
(1\,\iab)$~\cite{buschmann,schlaffer_spannowsky,sanz,azatov,andi}.
Because the logarithmic dependence in Eq.\eqref{eq:pt_log} is the same
for the 1-jet and 2-jet hard matrix elements, and because our
discussion in Sec.~\ref{sec:mass} shows that the same holds true even
for relatively low $p_{T,H} \gtrsim m_t$, we have to use multi-jet
merging to provide a reliable estimate of these effects. This also
implies that in an optimized combined analysis we should focus on
jet-inclusive observables like $p_{T,H}$ to combine different jet
multiplicities. As a side effect, the merged approach guarantees a
reliable description for the distributions over the full momentum
range, as we can see in Figs.~\ref{fig:pth_merged_LO}
and~\ref{fig:pth_merged_NLO}.\bigskip

%-------------------------------------------------------
\begin{figure}[t!]
 \includegraphics[width=0.4\textwidth]{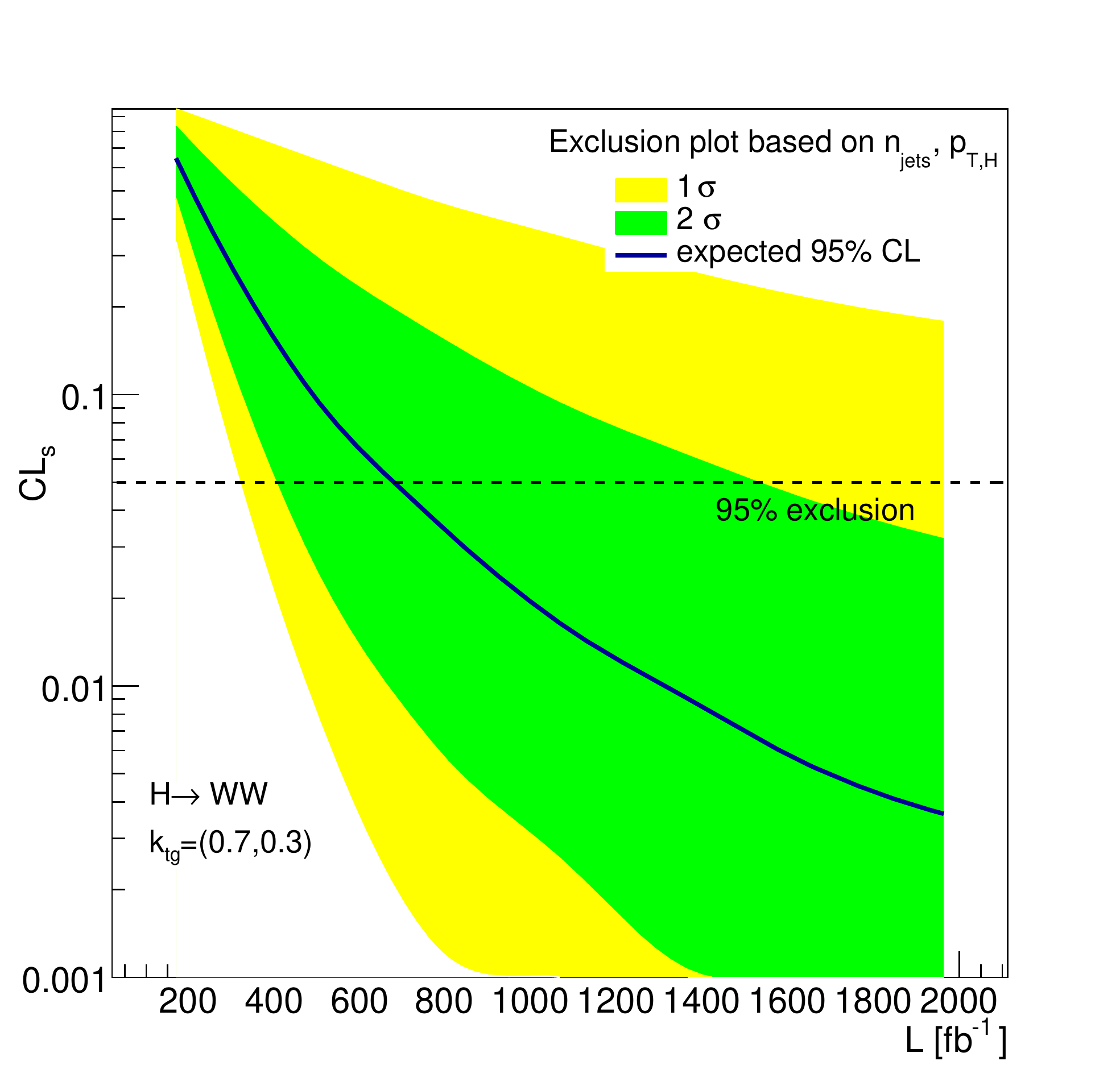}
\caption{Confidence level for separating the BSM hypotheses
  $\kappa_{t,g}=(0.7,0.3)$ from the Standard Model. We show results
  for $H\rightarrow WW$ decays based on the 2D distribution
  $(n_\text{jets},p_{T,H})$.}
\label{fig:cls_boosted} 
\end{figure}
%-------------------------------------------------------

In Fig.~\ref{fig:cuts1} we show two of the key distributions which
allow us to separate the Higgs signal from the different
background. The azimuthal angle between the two forward jets is a well
known probe for the Lorentz structure of the hard
process~\cite{phi_jj}.  Because it is only defined for at least two
additional hard jets, it is one of the key improvements of our merged
analysis over Higgs production with a single hard jet.  The
corresponding background rejection cuts are given in
Tab.~\ref{tab:cuts1}.  The signal events are generated with NLO
predictions for the 0-jet and 1-jet processes and LO precision for the
hard 2-jet process.  The $t\bar{t}$+jets background includes the NLO
prediction for the 0-jet bin and up to 3 hard jets at the LO
level. The QCD component from $WW$+jets production is loop induced and
it is generated with up to 1-jet with LO precision.  The electroweak
$WW$+jets component includes NLO corrections up to the 1-jet bin and
up to 3 hard jets at LO.  The analysis largely correspond to the $WW$
analysis proposed for the 2-jet channel in Ref.~\cite{buschmann}.\bigskip

In particular when we link different experimental measurements to the
same Lagrangian interpretation given by Eq.\eqref{eq:lagrangian} the
question arises how the experimental approaches compare. In
Fig.~\ref{fig:cls_boosted} we show an idealized projection of the
reach in the boosted Higgs analyses. Based on a 2-dimensional
$\text{CL}_s$ analysis of the number of jets and the transverse
momentum of the Higgs, $(n_\text{jets},p_{T,H})$, we estimate how much
luminosity would be required to rule out our BSM reference point given
in Eq.\eqref{eq:points}. We find that in the absence of systematic and
theoretical uncertainties it would take around $700~\ifb$ of data to
achieve a 95\% C.L exclusion. Even if we rely on the reference region
at $p_{T,H} \lesssim m_t$ to efficiently reduce the uncertainties, a
meaningful study of boosted Higgs production might require attobarn
integrated luminosities. On that time scale it appears unlikely that
such a detailed kinematic analysis will be able to compete with a
dedicated hypothesis test based on Higgs couplings and including
$t\bar{t}H$ production with the combined Higgs decays $H \to b\bar{b},
\tau \tau, \gamma \gamma$~\cite{tth,sfitter}.

%%%%%%%%%%%%%%%%%%%%%%%%%%%%%%%%%%%%%%%%%%%%%%%%%%
\section{Off-shell Higgs production}
\label{sec:offshell}

One of novel LHC measurement in 2014 is the Higgs width limit from
off-shell Higgs production~\cite{offshell,offshell_ex}. For example,
CMS published the first results on a rate measurement of $pp
\rightarrow Z^{(*)}Z^{(*)}\rightarrow$~4 leptons at high invariant
mass $m_{4\ell}$~\cite{offshell_ex}. This process relies on off-shell
contributions from $s$-channel Higgs production. This way, it carries
information on the Higgs couplings at different energy scales which
could, similarly to the boosted case, probe the energy dependence of
the higher-dimensional operators. In this spirit, we exploit the
off-shell Higgs regime of $ZZ$ production to probe the Higgs--top
sector of the Standard Model~\cite{taming,cacciapaglia}.\bigskip

%-------------------------------------------------------
\begin{figure}[t!]
 \includegraphics[width=0.95\textwidth]{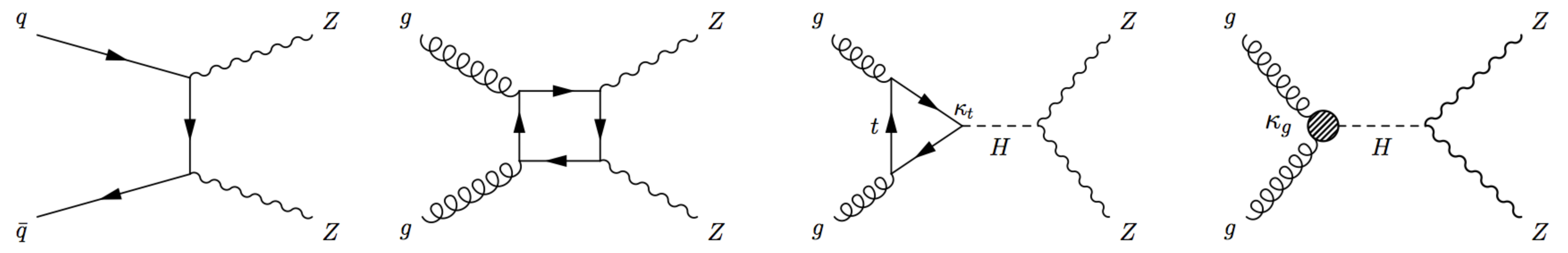}
\caption{Sample Feynman diagrams for the continuum background
  $q\bar{q}\,(gg)\rightarrow ZZ$ (left) and for the signal ${gg
    \rightarrow H \rightarrow ZZ}$ (right) with full top mass
  dependence and in the $m_t \rightarrow \infty$ approximation.}
\label{fig:m4l_diag} 
\end{figure}
%-------------------------------------------------------

The Higgs contribution to $Z$-pair production is generated via gluon
fusion through heavy quark loops. It faces two dominant backgrounds:
$q\bar{q}\rightarrow ZZ$ and $gg\rightarrow ZZ$. The $q\bar{q}$
component is generated already at the tree level and constitutes the
most important contribution. It is approximately one order of
magnitude larger than the gluon fusion part. On the other hand, the
gluon fusion contribution features an interference with the Higgs
signal in the off-shell $m_{ZZ}$ regime.  In Fig.~\ref{fig:m4l_diag}
we display a sample of the contributing Feynman diagrams to $ZZ$
production.

%-------------------------------------------------------
\begin{figure}[t]
 \includegraphics[width=0.45\textwidth]{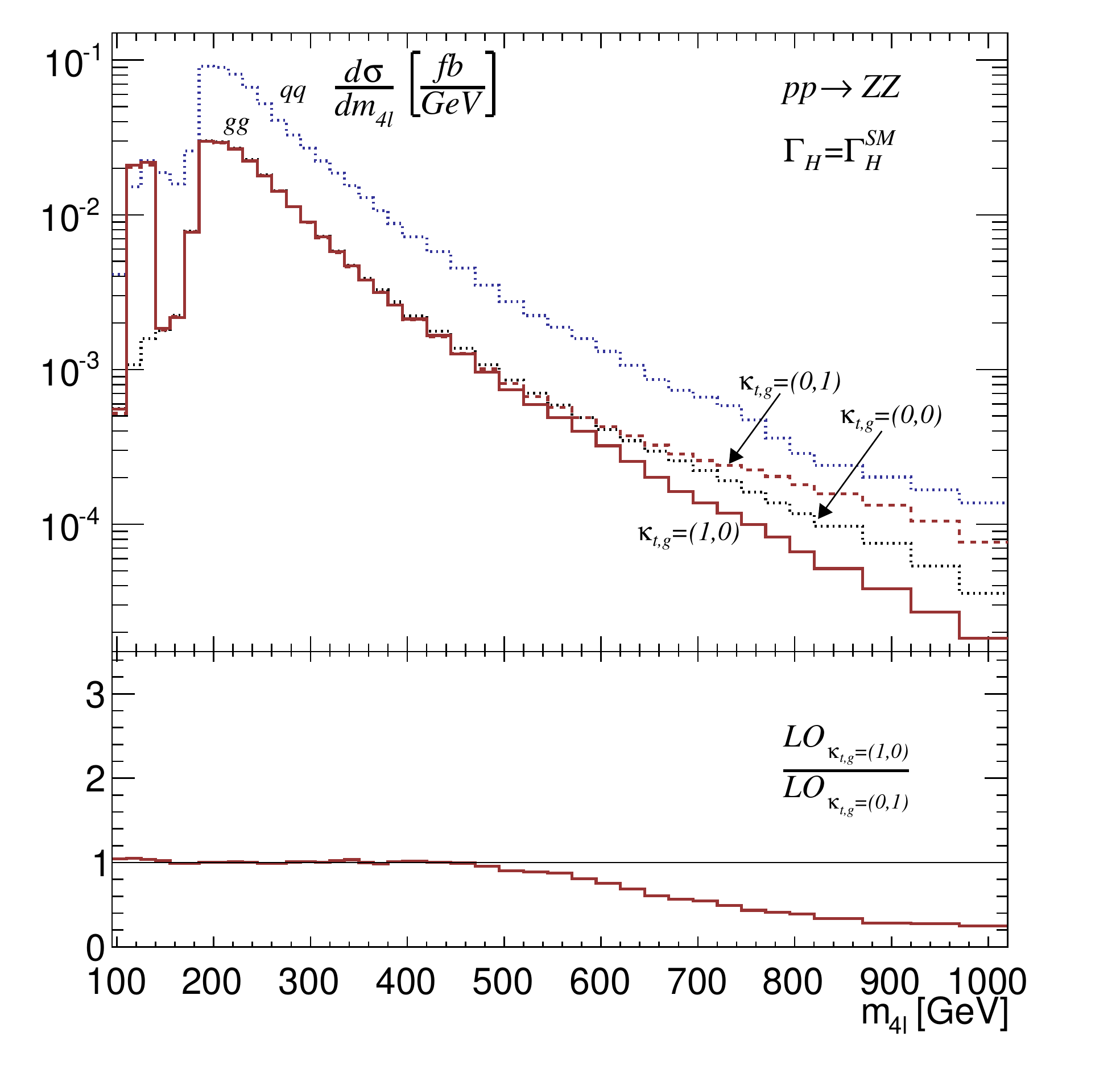}
 \includegraphics[width=0.45\textwidth]{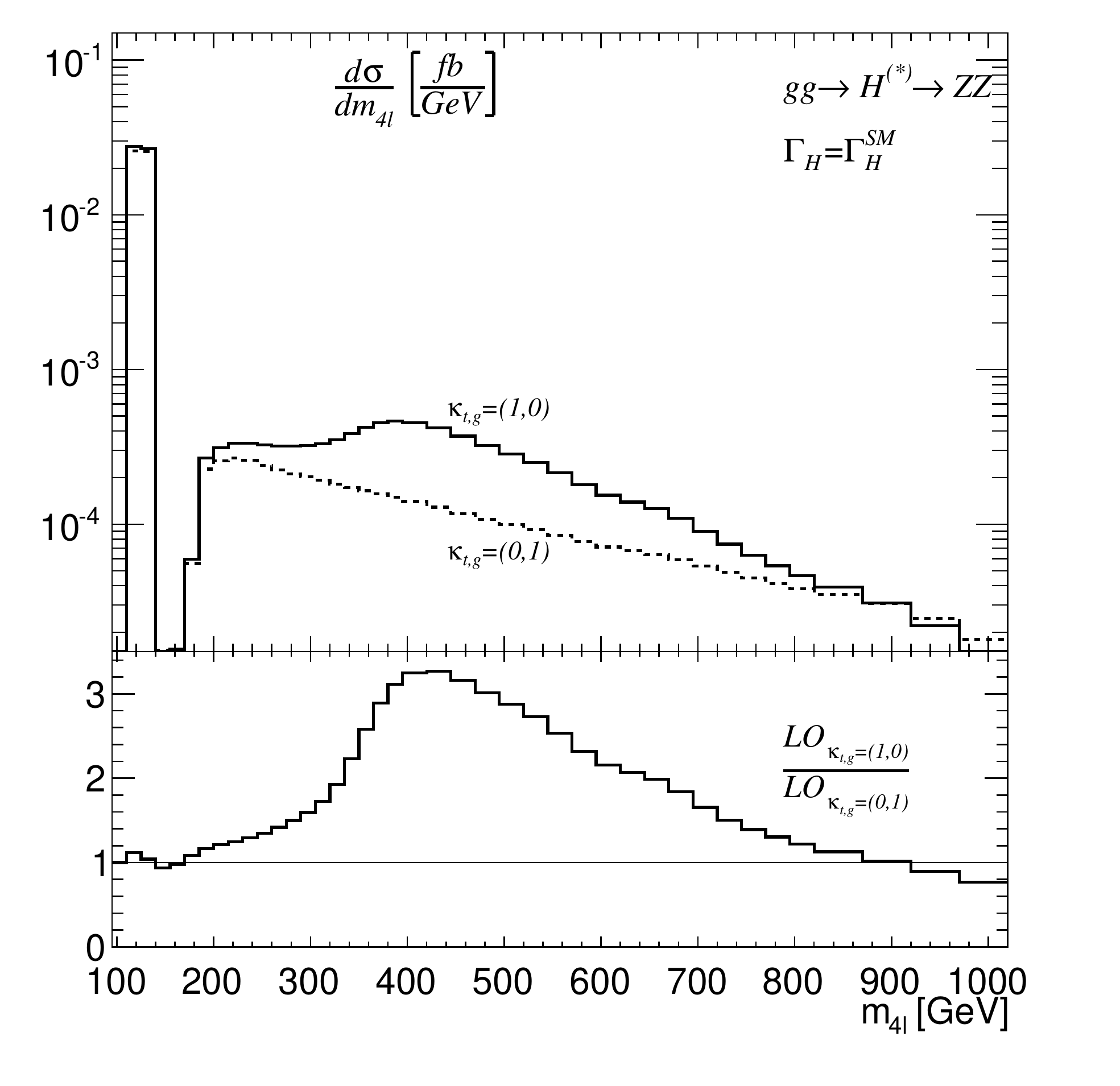}
\caption{$m_{4l}$ distributions for the $q\bar{q}(gg)\rightarrow ZZ$ (left
  panel) and $gg\rightarrow H^{(*)} \rightarrow ZZ$ (right panel) for the
  different signal hypothesis and the dominant background.  We assume
  the LHC at $\sqrt{S}=13$~TeV.}
\label{fig:m4l} 
\end{figure}
%-------------------------------------------------------

At high invariant mass $m_{ZZ}$ the Higgs decays mostly into
longitudinal gauge bosons. This means that the signal amplitude can be
understood from the longitudinal
components~\cite{glover_zz,taming,cacciapaglia}
\begin{alignat}{5}
\mat_t^{++00}
%&=
%\frac{m_t^2\left(m_{4\ell}^2-2m_Z^2 \right)}
%{m_Z^2\left( m_{4\ell}^2-m_H^2+i\Gamma_Hm_H \right)}
%\left[ -2+\left(m_{4\ell}^2-4m_t^2\right)C \right] \notag \\ 
%&=
%\frac{m_t^2}{m_Z^2} \; 
%\frac{m_{4\ell}^2-2m_Z^2}{m_{4\ell}^2-m_H^2+i\Gamma_Hm_H} \; 
%\left[ -2+ m_{4\ell}^2 \left(1-\frac{4m_t^2}{m_{4\ell}^2} \right)C \right] \not%ag \\
%&=
%\frac{m_t^2}{m_Z^2} \; 
%\frac{m_{4\ell}^2-2m_Z^2}{m_{4\ell}^2-m_H^2+i\Gamma_Hm_H} \; 
%\left[ -2- m_{4\ell}^2 \left(1-\frac{4m_t^2}{m_{4\ell}^2} \right) 2 \frac{f(\ta%u)}{m_{4\ell}^2} \right] \notag \\
&=
- 2\; \frac{m_{4\ell}^2-2m_Z^2}{m_Z^2} \; 
\frac{m_t^2}{m_{4\ell}^2-m_H^2+i\Gamma_Hm_H} \; 
\left[ 1 + \left(1-\frac{4m_t^2}{m_{4\ell}^2} \right) 
       f\left(\frac{4m_t^2}{m_{4\ell}^2}\right) \right] \; ,
\label{eq:m4l_full}
\end{alignat}
where $\Gamma_H$ is the Higgs boson width and
$f = - m_{4\ell}^2 C(m_{4\ell}^2;m_t,m_t,m_t)/2$ represents the dimensionless scalar
three-point function. This form corresponds to the on-shell case in
Eq.\eqref{eq:higgs_eff1}, replacing $\tau = 4m_t^2/m_H^2$ with its
off-shell analogue $4 m_t^2/m_{4\ell}^2$ and relying on the form
\begin{equation}
f(\tau)= - \frac{1}{4} \left( \log \dfrac{1+\sqrt{1-\tau}}
		                                     	            {1-\sqrt{1-\tau}}
	         		        - i\pi\right)^2
\end{equation}
for the scalar integral with $\tau < 1$.
In the low-energy limit far above the Higgs mass shell, 
$m_t \gg m_{4\ell} \gg m_H,m_Z$, the scalar integral scales like
$f \sim m_{4\ell}^2/(4 m_t^2)$ and gives the usual finite effective Higgs--gluon 
coupling $g_{ggH}$ defined in Eq.\eqref{eq:higgs_eff1}. Obviously, this assumption
is not correct once we include the actual mass values. Instead, for $\mat_t$ we
better assume $m_{4\ell} \gg m_t \gtrsim m_H,m_Z$, giving us
\begin{alignat}{5}
\mat_g^{++00}
&\approx
-\frac{m_{4\ell}^2}{2m_Z^2} \qqqquad 
&&\text{with} \; m_t \gg m_{4\ell} \gg m_H,m_Z \notag \\
\mat_t^{++00}
&\approx
+\frac{m_t^2}{2m_Z^2}
\log^2 \frac{m_{4\ell}^2}{m_t^2}\qqqquad 
&&\text{with} \; m_{4\ell} \gg m_t \gtrsim m_H,m_Z \notag \\
\mat_c^{++00}
&\approx
-\frac{m_t^2}
{2m_Z^2}\log^2 \frac{m_{4\ell}^2}{m_t^2} 
&&\text{with} \; m_{4\ell} \gg m_t \gtrsim m_Z \; .
\label{eq:m4l_tgc}
\end{alignat}
In the proper limit a logarithmic dependence on $m_{4\ell}/m_t$
develops far above the Higgs mass shell. It is very similar to the
transverse momentum dependence in the boosted regime, as seen in 
Eq.\eqref{eq:pt_log}. The ultraviolet logarithm cancels between the
correct Higgs amplitude and the continuum, ensuring the proper
ultraviolet behavior of the full amplitude.  Most importantly, there
appears a sign difference between the full top mass dependence and the
low-energy limit.  For the interference pattern with the continuum
process $gg\rightarrow ZZ$ the full top mass dependence predicts a
destructive interference whereas in the low-energy limit the
interference far above mass shell should be constructive.\bigskip

Following the parametrization in Eqs.~\eqref{eq:lagrangian}
and~\eqref{eq:amplitude_boost} and including the interference with the
continuum background arising from the box diagrams, we can write the
gluon-induced amplitude $gg\to ZZ$ as
\begin{alignat}{5}
\mat_{ZZ} & =
\kappa_t \mat_t + \kappa_g\mat_g + \mat_c  \; .
\label{eq:amplitude_offshell}
\end{alignat}
Correspondingly, the differential cross section can be expressed as
\begin{alignat}{5}
\frac{d\sigma}{dm_{4\ell}} & =
\frac{d\sigma_c}{dm_{4\ell}}+
\kappa_t \frac{d\sigma_{tc}}{dm_{4\ell}}+
\kappa_g \frac{d\sigma_{gc}}{dm_{4\ell}}+
\kappa_t^2 \frac{d\sigma_{tt}}{dm_{4\ell}}+
\kappa_t \kappa_g \frac{d\sigma_{tg}}{dm_{4\ell}}+
\kappa_g^2 \frac{d\sigma_{gg}}{dm_{4\ell}}
\;.
\label{eq:distribution_offshell}
\end{alignat}
Using this parametrization, we can access each of the
different components by switching on and off the coefficients
$\kappa_t$ and $\kappa_g$.\bigskip

The gluon--initiated and quark--initiated $pp \to e^+e^-\mu^+\mu^-$
signal and background events are generated with
\textsc{MCFM-6.8}~\cite{mcfm}, respectively at LO and NLO.  We modify
the original \textsc{MCFM} code to separately access all components
defined in Eq.\eqref{eq:distribution_offshell}.  All our results
follow the CMS cut-flow analysis~\cite{offshell_ex}
\begin{alignat}{5}
p_{T,\mu} &>5~\gev      \qquad & |\eta_\mu|&<2.4 \notag \\
p_{T,e} &>7~\gev        \qquad & |\eta_e|&<2.5 \notag \\
m_{\ell\ell'} &>4~\gev  \qquad & m_{4\ell}&>100~\gev \;.
\label{eq:m4l_cuts}
\end{alignat}
For the decay leptons we require transverse momenta above $20$
$(10)~\gev$ for the leading (sub-leading) lepton and invariant masses
of ${40<m_{\ell\ell}<120~\gev}$ (${12<m_{\ell\ell}<120~\gev}$) for the
leading (sub-leading) same-flavor lepton pair.  We take the
renormalization and factorization scales to be $m_{4\ell}/2$ and use
the PDF set CTEQ6L1~\cite{cteq}.

%-------------------------------------------------------
\begin{figure}[t!]
 \includegraphics[width=0.32\textwidth]{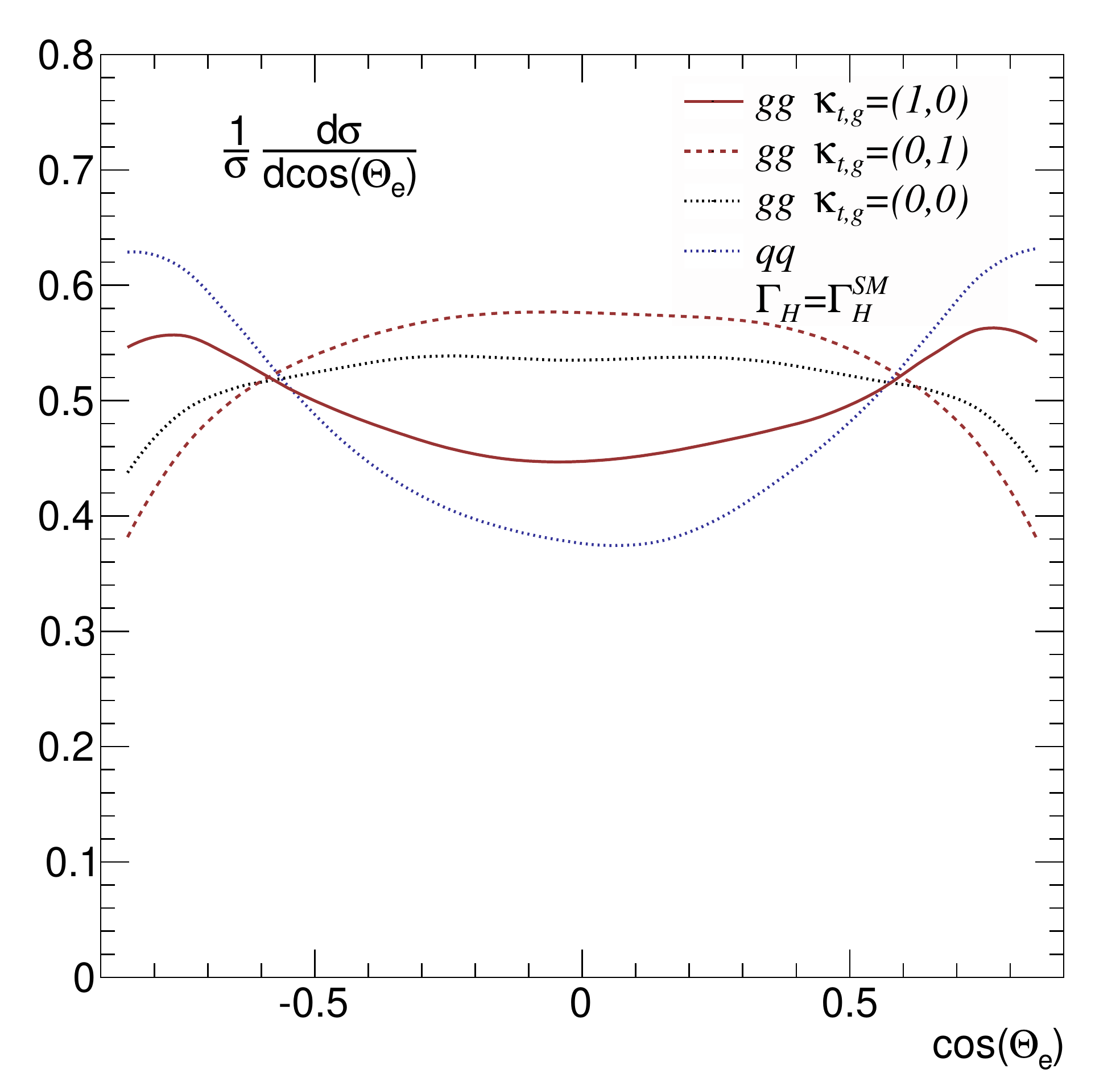}
 \hspace*{0.1\textwidth}
 \includegraphics[width=0.32\textwidth]{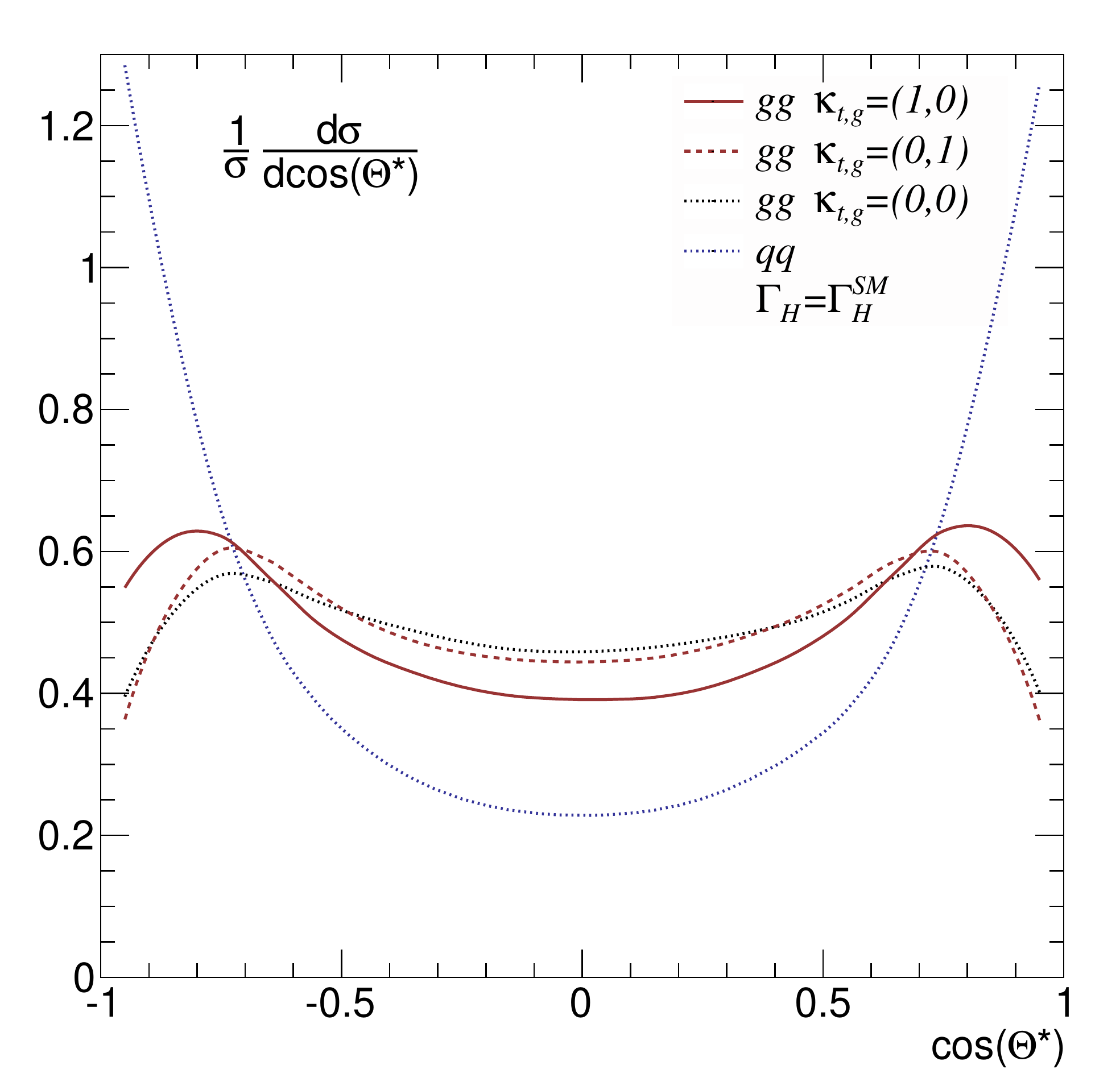} \\
 \includegraphics[width=0.32\textwidth]{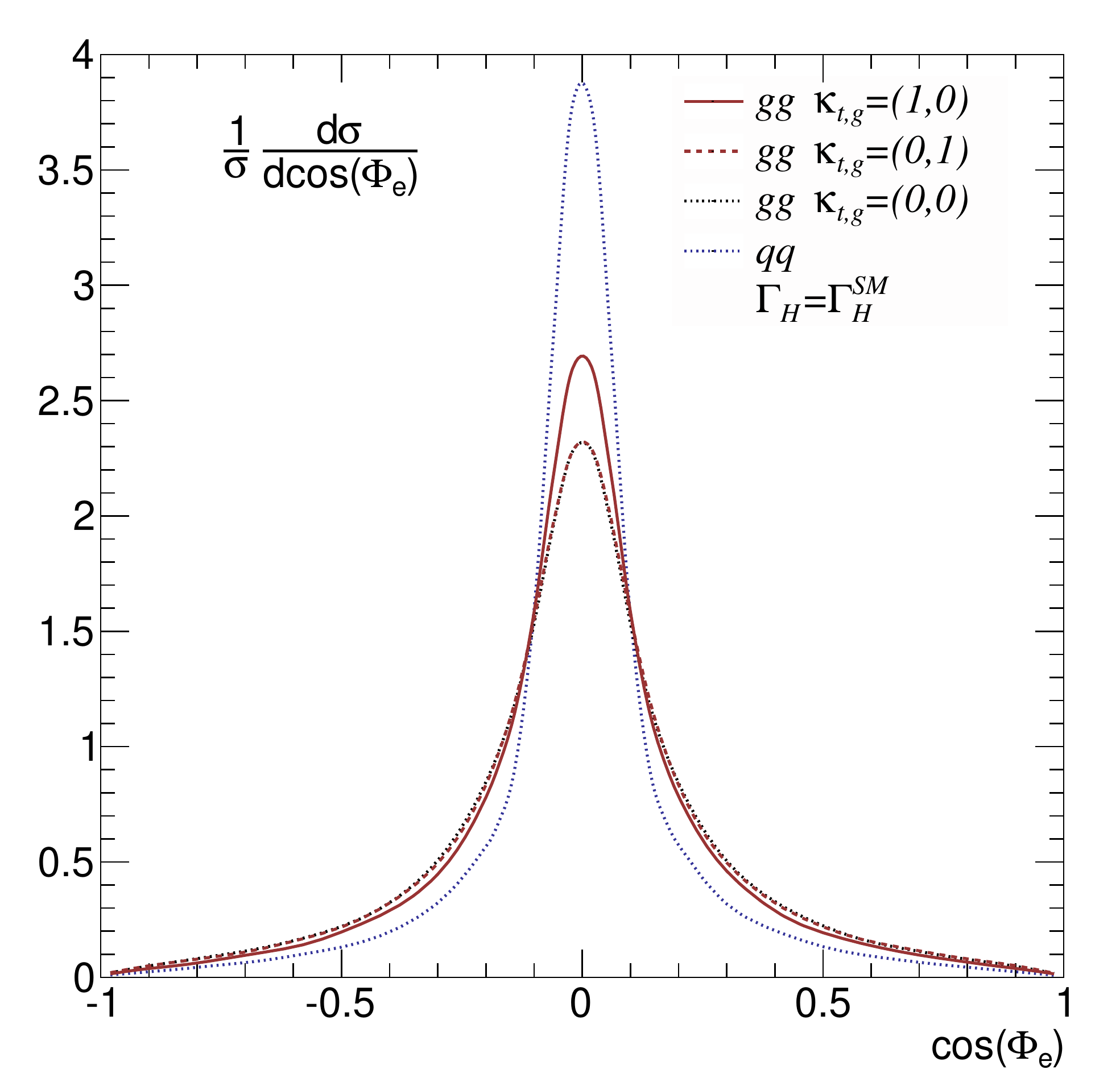}
 \hspace*{0.1\textwidth}
 \includegraphics[width=0.32\textwidth]{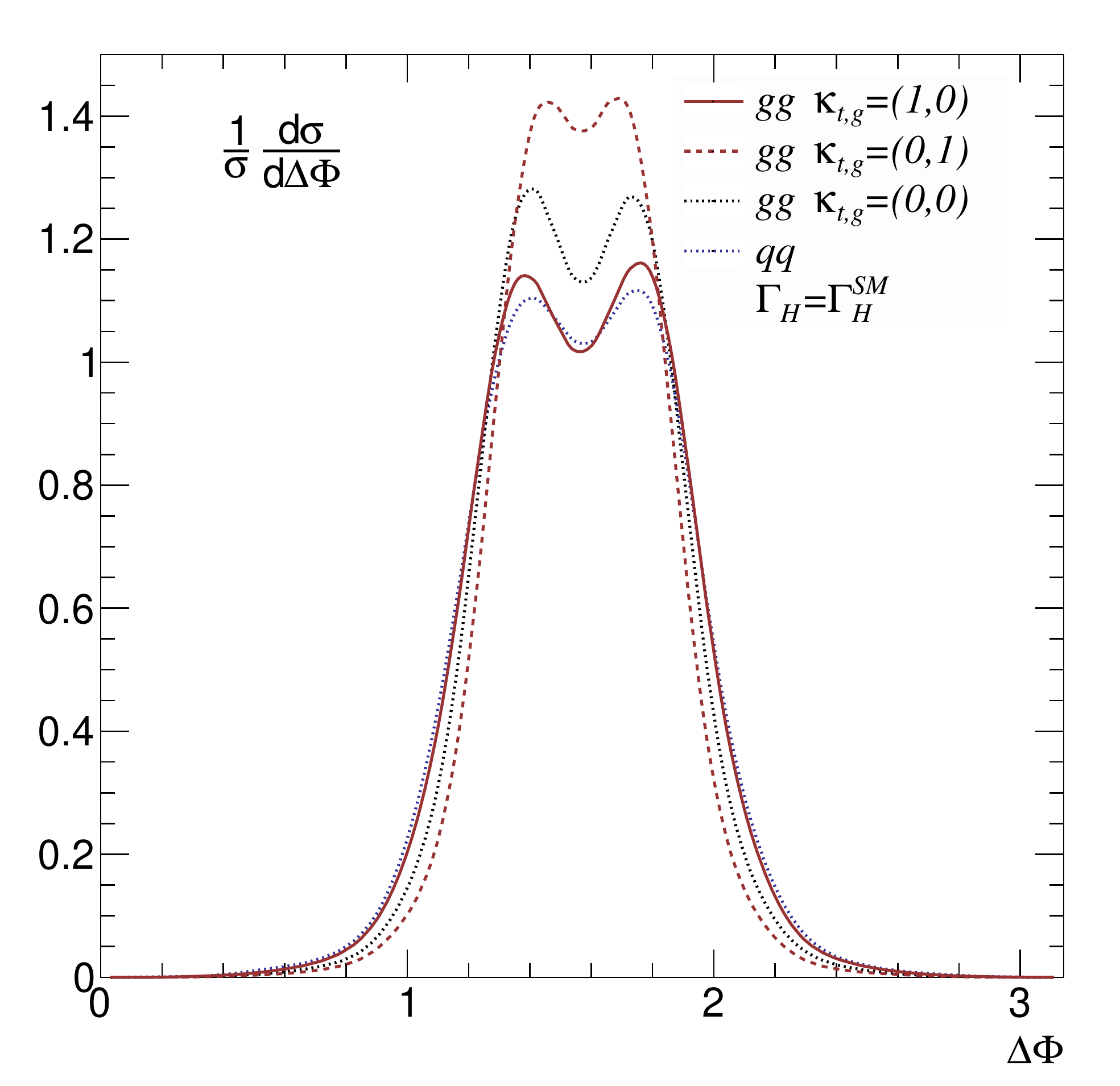}
\caption{Normalized distributions for the angular correlations in the
  $gg\rightarrow ZZ$ assuming different signal hypothesis
  $\kappa_{t,g}$. We also show the quark-induced background $q\bar{q}
  \rightarrow ZZ$ for $m_{4\ell}>600~\gev$.  We assume the LHC at
  $\sqrt{S}=13$~TeV.}
\label{fig:angles} 
\end{figure}
%-------------------------------------------------------

The Higgs production process $gg\rightarrow H^{(*)} \rightarrow ZZ$
has been computed to very high precision. This includes NNLO and NNLL
contributions for inclusive production and at NNLO for the $m_{ZZ}$
distribution~\cite{passarino_cat}. On the other hand, the continuum
production $gg\rightarrow ZZ$ is only known to leading order. However,
it was recently demonstrated that the NLO and NNLO effects for the
signal and the signal-background interference contributions have a
very similar perturbative QCD
enhancement~\cite{melnikov_interference}. Therefore, we could include
these QCD effects by assuming $K$-factors for the signal,
signal-background interference and background contributions. In this
study we include a differential NLO $K$-factor only.\bigskip

In Fig.~\ref{fig:m4l} we present the distributions for the different
signal hypotheses defined in Eq.\eqref{eq:amplitude_offshell} along
with the quark--induced background. In the bottom panel we show the
ratio between the top mass dependent coupling $\kappa_{t,g}=(1,0)$ and
the low-energy form $\kappa_{t,g}=(0,1)$.  As expected from
Eq.\eqref{eq:m4l_tgc}, in the full process $gg\rightarrow ZZ$ we
observe an enhancement in the tail for the low-energy limit and a
suppression for the full top mass result. In the right panel we focus
on Higgs production $gg\rightarrow H^{(*)} \rightarrow ZZ$ and neglect
the dominant continuum interference. For this process we observe three
threshold effects which largely control the $m_{4\ell}$ distribution
profile: first, we see the on-shell Higgs production peak around $m_H
\approx 125~\gev$; second, there is a threshold at $2m_Z \approx
180~\gev$ which leads to an abrupt enhancement in the cross section
without which the off-shell measurements would not be
possible~\cite{offshell_ex}; third, a peak appears when both tops in
the loop-induced coupling are simultaneously on-shell, $2m_t \approx
350~\gev$. This last threshold is only present when we include the
full top mass dependence.  This leads to a clear enhancement from the
absorptive top threshold with respect to the effective coupling
approach, tantamount to a maximum correction factor of
$\mathcal{O}(3)$ shown in the bottom-right panel of
Fig.~\ref{fig:m4l}.\bigskip

While there are no spin correlations between production and decay in
the Higgs amplitudes ${|\mat_{t,g}|^2}$, they appear through the
interference with the continuum background $\mat_c$. This way, the
Cabibbo--Maksymowicz--Dell'Aquila--Nelson angles~\cite{nelson} help to
extract top mass effects and allow us to further suppress the
background in the off-shell regime.  The set of five angles completely
capture the kinematics of the decay $H\to ZZ\to 4\ell$, including the
correlation with the incoming protons.  The momenta from the Higgs
decay are given by
\begin{alignat}{5}
p_H = p_{Z_e}+p_{Z_\mu} 
\qqquad 
p_{Z_e} = p_{e^-}+p_{e^+} 
\qqquad 
p_{Z_\mu} = p_{\mu^-}+p_{\mu^+} \; .
\end{alignat}
For each of these momenta and the beam direction we define unit
three-momenta $\hat{p}_i$ in the Higgs rest frame and in the two
$Z_{e,\mu}$ rest frames
\begin{alignat}{5}
&\cos \theta_e =   
  \hat{p}_{e^-} \cdot\hat{p}_{Z_\mu} \Big|_{Z_e} 
&\qquad 
&\cos \theta_\mu =   
  \hat{p}_{\mu^-} \cdot\hat{p}_{Z_e} \Big|_{Z_\mu} 
\qqqquad 
\cos \theta^* = 
  \hat{p}_{Z_e} \cdot \hat{p}_\text{beam} \Big|_H  \notag \\
&\cos \phi_e = 
  (\hat{p}_\text{beam} \times \hat{p}_{Z_\mu}) \cdot (\hat{p}_{Z_\mu} \times \hat{p}_{e^-}) \Big|_{Z_e} 
  &\qquad
&\cos \Delta \phi = 
  (\hat{p}_{e^-} \times \hat{p}_{e^+}) \cdot (\hat{p}_{\mu^-} \times \hat{p}_{\mu^+}) \Big|_H \; .
\label{eq:angles_zz}
\end{alignat}
In Fig.~\ref{fig:angles} we present the the normalized angular
distributions. Notice that the main sensitivity at high $m_{4\ell}$
appears in the two polarization angles $\cos\theta_\ell$.\bigskip

%-------------------------------------------------------
\begin{figure}[t!]
 \includegraphics[width=0.4\textwidth]{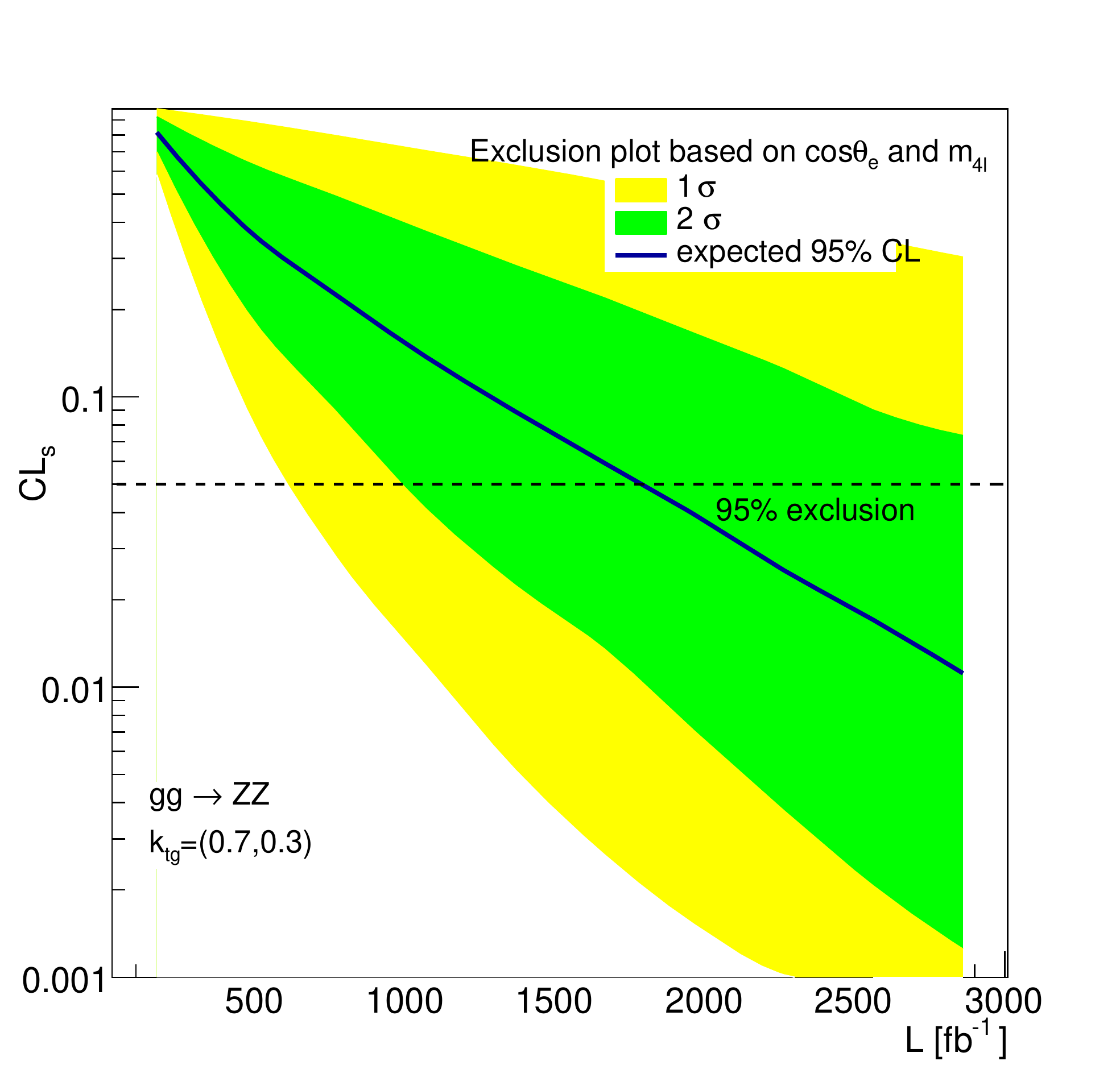}
\caption{Confidence level for separating the BSM hypotheses
  $\kappa_{t,g}=(0.7,0.3)$ from the Standard Model. We show results
  for $gg\rightarrow ZZ$ decays based on the 2D distribution $(\cos
  \theta_e,m_{4\ell})$.}
\label{fig:cls_offshell} 
\end{figure}
%-------------------------------------------------------

To roughly compare the potential of the boosted Higgs production and
off-shell Higgs production in probing the effective Higgs--gluon
coupling we analyze the information from the different $m_{4\ell}$
bins using the $\text{CL}_s$ method. To enhance the signal sensitivity
we use the angular correlations in two ways. First, we suppress the
$q\bar{q}\rightarrow ZZ$ background by requiring that $|\cos \Theta^*|<0.7$.
Second, we include the leading angle in a 2-dimensional
$\text{CL}_s$ analysis of $(\cos \theta_e,m_{4\ell})$.  In
Fig.~\ref{fig:cls_offshell} we see that even using an overly
optimistic setup, where we neglect systematic as well as theory
uncertainties, the off-shell channel will only allow for an exclusion
of our BSM hypothesis defined in Eq.\eqref{eq:points} based on a few
inverse attobarns of integrated luminosity. Compared to the reach of
the idealized results from boosted Higgs production shown in
Fig.~\ref{fig:cls_boosted} this is not promising. Moreover,
measuring the top Yukawa using $t\bar{t}H$ production and the
combined Higgs decays $H \to b\bar{b}, \tau \tau, \gamma \gamma$ will
easily constrain a modified structure of the Higgs--gluon coupling
beyond this level~\cite{sfitter}.

%%%%%%%%%%%%%%%%%%%%%%%%%%%%%%%%%%%%%%%%%%%%%%%%%%
\section{Width measurement}
\label{sec:width}

%-------------------------------------------------------
\begin{figure}[b!]
 \includegraphics[width=0.45\textwidth]{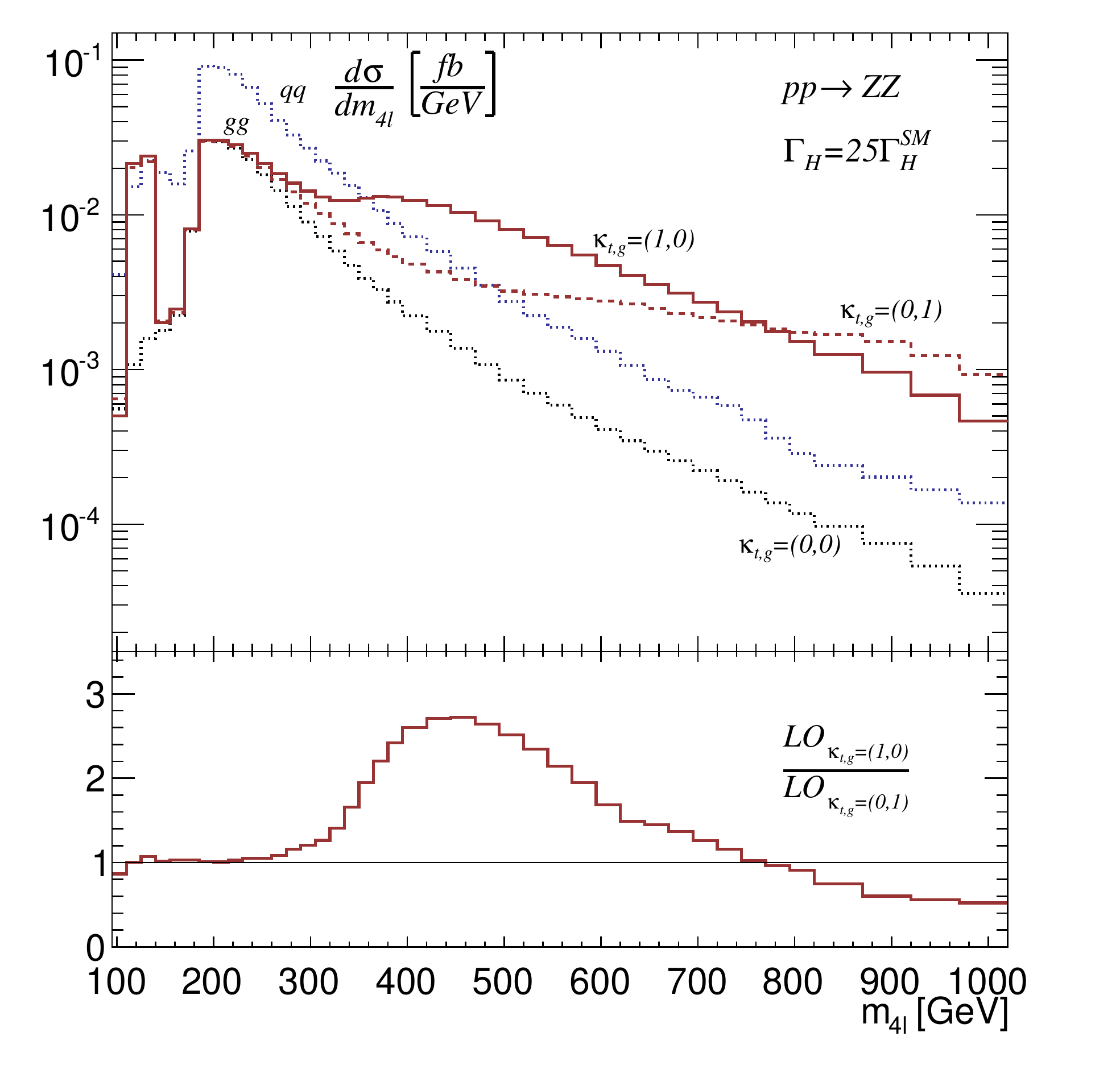}
 \includegraphics[width=0.45\textwidth]{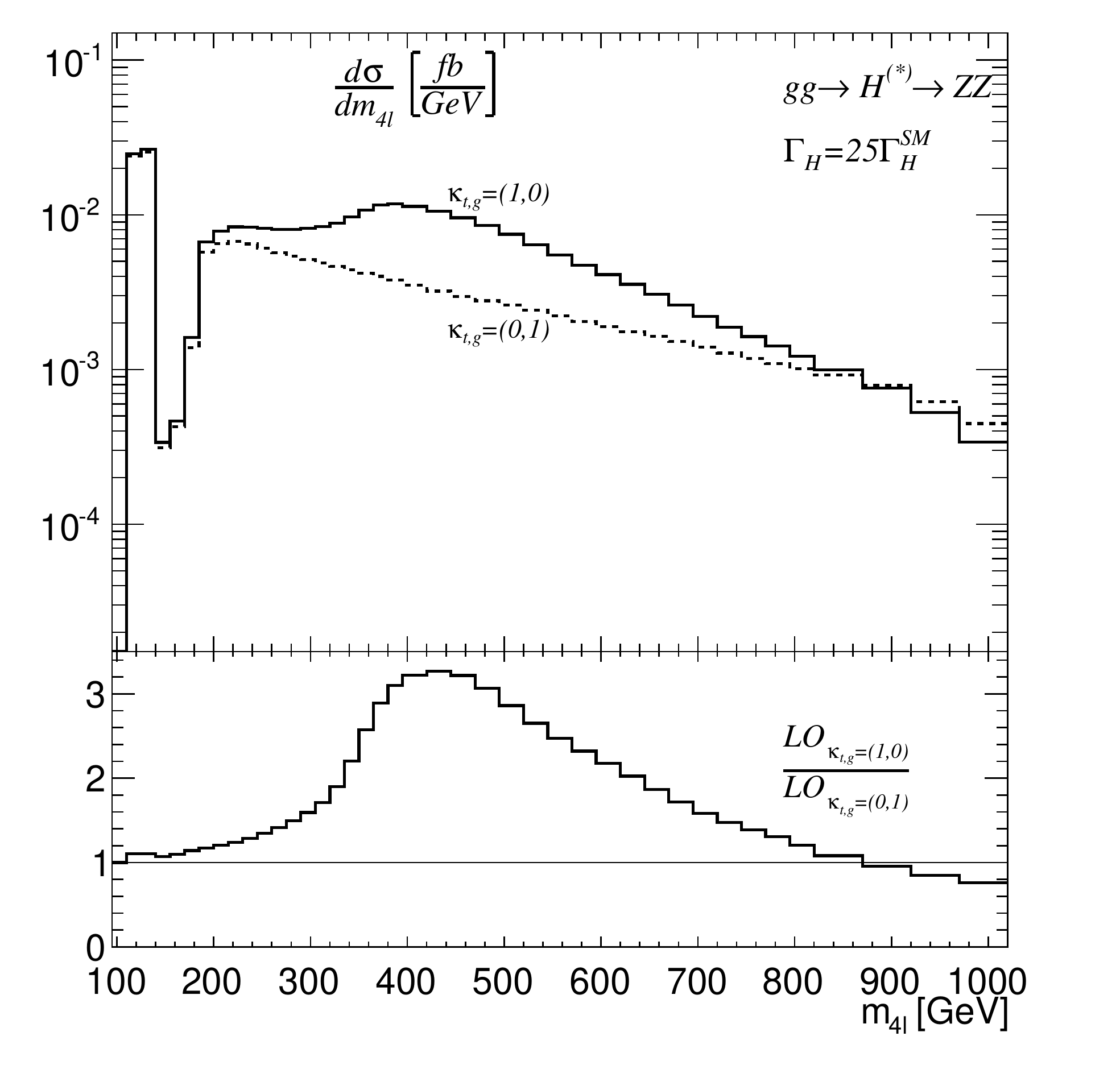}
\caption{$m_{4l}$ distributions for the $q\bar{q}(gg)\rightarrow ZZ$ (left panel) and
  $gg\rightarrow H^{(*)} \rightarrow ZZ$ (right panel) for the different signal
  hypothesis and the dominant background. For illustration purposes we
  assume an increased Higgs $\Gamma_H= 25\Gamma_{SM}$.}
\label{fig:m4l_width} 
\end{figure}
%-------------------------------------------------------

If combined with on-shell rate measurements, the additional off-shell
rate measurements described in Sec.~\ref{sec:offshell} can be used to
test the assumptions on the recently proposed Higgs width
measurement~\cite{offshell,offshell_ex}.  Fundamentally, this proposal
aims to break the degeneracy present on the on-shell Higgs couplings
analysis, namely
\begin{alignat}{5}
\sigma^{\text{on-shell}}_{i \rightarrow H \rightarrow f} 
\propto 
\frac{g^2_{i}(m_H) g^2_{f}(m_H)}{\Gamma_H} 
= \text{constant for}
\quad 
g_{i,f}(m_H) \to \xi g^{SM}_{i,f}(m_H) 
\quad
\Gamma_H \to \xi^4 \Gamma_{H} \;,
\label{eq:rescaling}
\end{alignat}
For example in the \textsc{SFitter} approach this degeneracy is broken
through the assumption that the total Higgs width be the sum of all
observed particle Higgs widths~\cite{sfitter}. Alternatively, in
\textsc{HiggsSignals}~\cite{higgs_signals} the unitarization of the
$WW$ scattering amplitude is turned into an upper limit on the Higgs
couplings to break this degeneracy.  Using off-shell rates we exploit
the subleading dependence on the Higgs width $\Gamma_H$ in the
off-shell regime, ${\sigma^\text{off-shell}_{i \rightarrow H^*
    \rightarrow f} \propto g^2_{i}(\sqrt{\hat{s}})
  g^2_{f}(\sqrt{\hat{s}})}$.  While it is an interesting idea, it is
clearly not a model independent width
measurement~\cite{offshell_model}. The model dependence immediately
follows from Eqs.~\eqref{eq:higgs_eff1} and~\eqref{eq:m4l_full}, which
reflect the non-trivial momentum dependence of the loop--induced
Higgs--gluon coupling. The effective coupling $g_{ggH}$ can only be
treated as a constant number if all external particles are on their
respective mass shells. For tree-level Higgs couplings we could assume
a weak, well-defined logarithmic running, while for such a
loop--induced couplings the momentum structure can be anything. For
instance, in our framework the
$\kappa_t$ and $\kappa_g$ corrections to the on-shell rate factorize. In contrast,
in the off-shell regime we obtain a non-trivial dependence,
\begin{alignat}{5}
\sigma^\text{on-shell}_{gg \rightarrow H \rightarrow ZZ} 
&\propto 
(\kappa_t+\kappa_g)^2 \; \frac{g^2_{ggH}(m_H) g^2_{HZZ}(m_H)}{\Gamma_H} 
\notag \\
 \sigma^{\text{off-shell}}_{gg \rightarrow H^* \rightarrow ZZ} 
&\propto \left( \kappa_t g_{ggH}(m_{4\ell})+ \kappa_g g_{ggH}(m_H) \right)^2 \; g^2_{HZZ}(m_{4\ell})\;.
\label{eq:offshell}
\end{alignat}
If we assume a scaling factor $\xi^4=25$, the Higgs width will vary by
$\Gamma=25\Gamma_H$ while the on-shell signal strength will still be
at $\mu^\text{on-shell}=1$.  In Fig.~\ref{fig:m4l_width} we display
the outcome of this parameter changing. We observe that the on-shell
peak still has the same size as the SM one, as expected from
Eq.\eqref{eq:offshell}. However, the interference between the Higgs
diagram $gg \to H^* \to ZZ$ and the continuum process $gg \to ZZ$
shows a very different behavior as a function of $m_{4 \ell}$, also
shown in Eq.\eqref{eq:offshell}.

This comparison of a top--induced Higgs--gluon coupling with the full
top mass dependence and in the low-energy limit might be the easiest
way of observing the limitation of model independent Higgs width
measurements at the LHC.  We need to assume that the quantum structure
of the effective top--Higgs coupling is exactly as in the Standard
Model $\kappa_{t,g}=(1,0)$ to interpret the current off-shell rate
measurement as a measurement of the Higgs width. Additional states
increasing the Higgs width cannot contribute to the Higgs--gluon
coupling without significantly changing the relation between on-shell
and off-shell rates.

%%%%%%%%%%%%%%%%%%%%%%%%%%%%%%%%%%%%%%%%%%%%%%%%%%
\section{Summary}
\label{sec:summary}
 
Top mass effects in Higgs production in gluon fusion can be linked to
experimentally relevant changes in some of the fundamental
distributions, namely the Higgs transverse momentum and reconstructed
mass of the Higgs decay products. We have introduced a
state-of-the-art simulation tool based on next-to-leading order
multi-jet merging. Our results show that the top mass effects
factorize for each number of hard jets to leading order and
next-to-leading order. This approach allows us to combine Higgs
production in association with one and with two hard jets to optimally
probe the structure of the Higgs--gluon coupling. Once we include two
jets an experimental analysis based on Higgs decays to $WW$ pairs
might well be the most promising search channel.\bigskip

Off-shell production and boosted Higgs production can be interpreted
in the same physics framework. While keeping the total Higgs
production rate constant we can vary the top Yukawa coupling and an
additional dimension-6 Higgs--gluon coupling and search for deviations
in key distributions. For a benchmark point with a 30\% reduction of
the top Yukawa coupling compensated by a dimension-6 contribution from
physics beyond the Standard Model we estimate the required luminosity
for a 95\% C.L exclusion in both channels. In the absence of
systematic and theory uncertainties the boosted Higgs channel will
require luminosities close to $1~\iab$. Off-shell Higgs production
will only rule out the same benchmark value with close to $2~\iab$ of
integrated luminosity. These numbers can be compared to a standard
Higgs coupling analysis: including a conservative estimate of
systematic and theoretical uncertainties a coupling extraction of 7
independent Higgs couplings can measure the top Yukawa with a
precision of 12\% based on $3~\iab$ of LHC data~\cite{sfitter_future}.\bigskip

Finally, we study the top mass dependence of the Higgs--gluon coupling
and its impact on the proposed direct Higgs width measurement. It
turns out that the width measurement essentially requires the
knowledge of the quantum structure of the Higgs--gluon
coupling. Additional states responsible for unobserved Higgs decays
cannot be linked to this loop--induced interaction.

\newpage
%%%%%%%%%%%%%%%%%%%%%%%%%%%%%%%%%%%%%%%%%%%%%%%%
  
\end{document}

%\bibitem{stops}
%  A.~Arvanitaki and G.~Villadoro,
%  %``A Non Standard Model Higgs at the LHC as a Sign of Naturalness,''
%  JHEP {\bf 1202}, 144 (2012).
%  %[arXiv:1112.4835 [hep-ph]].
%  %%CITATION = ARXIV:1112.4835;%%
%  %84 citations counted in INSPIRE as of 28 Jul 2014